\documentclass[journal]{IEEEtran}

\usepackage{lineno, hyperref, amsmath, graphicx, subcaption}
\usepackage[dvipsnames]{xcolor}

\begin{document}
\title{Improved Lite Audio-Visual Speech Enhancement}

\author{Shang-Yi Chuang,
        Hsin-Min Wang,~\IEEEmembership{Senior Member,~IEEE,}
        Yu Tsao,~\IEEEmembership{Senior Member,~IEEE}% <-this % stops a space
\thanks{Shang-Yi Chuang and Yu Tsao are with Research Center for Information Technology Innovation, Academia Sinica, Taipei, Taiwan, corresponding e-mail: (yu.tsao@citi.sinica.edu.tw).}
\thanks{Hsin-Min Wang is with Institute of Information Science, Academia Sinica, Taipei, Taiwan.}
}

\markboth{}
{Shell \MakeLowercase{\textit{et al.}}: Bare Demo of IEEEtran.cls for IEEE Journals}

\maketitle

\begin{abstract}
Numerous studies have investigated the effectiveness of audio-visual multimodal learning for speech enhancement (AVSE) tasks, seeking a solution that uses visual data as auxiliary and complementary input to reduce the noise of noisy speech signals. Recently, we proposed a lite audio-visual speech enhancement (LAVSE) algorithm for a car-driving scenario. Compared to conventional AVSE systems, LAVSE requires less online computation and to some extent solves the user privacy problem on facial data. In this study, we extend LAVSE to improve its ability to address three practical issues often encountered in implementing AVSE systems, namely, the additional cost of processing visual data, audio-visual asynchronization, and low-quality visual data. The proposed system is termed improved LAVSE (iLAVSE), which uses a convolutional recurrent neural network architecture as the core AVSE model. We evaluate iLAVSE on the Taiwan Mandarin speech with video dataset. Experimental results confirm that compared to conventional AVSE systems, iLAVSE can effectively overcome the aforementioned three practical issues and can improve enhancement performance. The results also confirm that iLAVSE is suitable for real-world scenarios, where high-quality audio-visual sensors may not always be available.
\end{abstract}

\begin{IEEEkeywords}
speech enhancement, audio-visual, data compression, asynchronous multimodal learning, low-quality data
\end{IEEEkeywords}

\IEEEpeerreviewmaketitle

\section{Introduction}

\IEEEPARstart{S}{peech} is the most natural and convenient means for human-human and human-machine communications. In recent years, various speech-related applications have been developed and have facilitated our daily lives. For most of these applications, however, the performance may be affected by acoustic distortions, which may lower the quality of the input speech. These acoustic distortions may come from different sources, such as recording sensors, background noise, and reverberations. To alleviate the distortion issue, many approaches have been proposed, and speech enhancement (SE) is one of them. The goal of SE is to enhance low-quality speech signals to improve quality and intelligibility. SE systems have been widely used as front-end processes in automatic speech recognition (ASR) \cite{el2007evaluation, li2015robust, vincent2018audio}, speaker recognition \cite{li2011comparative}, speech coding \cite{li2011two}, hearing aids \cite{levit2001noise, venema2006compression, healy2019deep}, and cochlear implants \cite{chen2015evaluation, lai2016deep} to improve the performance of target tasks. 

Traditional SE methods are generally designed based on the properties of speech and noise signals. A class of approaches estimates the statistics of speech and noise signals to design a gain/filter function, which is then used to suppress the noise components in noisy speech. Notable examples belonging to this class include the Wiener filter \cite{scalart1996speech, chen2008fundamentals} and its extensions \cite{hansler2006topics}, such as the minimum mean square error spectral estimator \cite{makhoul1975linear, quatieri1992shape}, maximum a posteriori spectral amplitude estimator \cite{lotter2005speech, suhadi2010data}, and maximum likelihood spectral amplitude estimator \cite{mcaulay1980speech, kjems2012maximum}. Another class of approaches considers the temporal properties or data distributions of speech and noise signals. Notable examples include harmonic models \cite{frazier1976enhancement}, linear prediction models \cite{atal1979predictive, ephraim1992statistical}, hidden Markov models \cite{rabiner1986introduction}, singular value decomposition \cite{hu2002subspace}, and Karhunen-Loeve transform \cite{rezayee2001adaptive}. In recent years, numerous machine-learning-based SE methods have been proposed. These approaches generally learn a model from training data in a data-driven manner. Then, the trained model is used to convert the noisy speech signals into the clean speech signals. Notable machine-learning-based SE methods include compressive sensing \cite{wang2016compressive}, sparse coding \cite{eggert2004sparse, chin2017speaker}, non-negative matrix factorization \cite{mohammadiha2013supervised}, and robust principal component analysis \cite{candes2011robust, huang2012singing}.

More recently, deep learning (DL) has became a popular and effective machine learning algorithm \cite{ronneberger2015u, he2016deep, vaswani2017attention} and has brought significant progress in the SE field \cite{zhang2016deep, pascual2017segan, michelsanti2017conditional, luo2018tasnet, zhang2019research, xu2019spatial, kim2020t, hu2020dccrn, yang2020characterizing}. Based on the deep structure, an effective representation of the noisy input signal can be extracted and used to reconstruct a clean signal \cite{williamson2015complex, wang2018supervised, zheng2018phase, plantinga2020phonetic, wang2020cross, qi2020exploring, carbajal2020joint}. Various DL-based model structures, including deep denoising autoencoders \cite{lu2013speech, xia2014wiener}, fully connected neural networks \cite{liu2014experiments, xu2015regression, kolbk2017speech}, convolutional neural networks (CNNs) \cite{fu2017complex, pandey2019new}, recurrent neural networks (RNNs), and long short-term memory (LSTM) \cite{campolucci1999line, weninger2014single, erdogan2015phase, chen2015speech, weninger2015speech, sun2017multiple}, have been used as the core model of an SE system and have been proven to provide better performance than traditional statistical and machine-learning methods. Another well-known advantage of DL models is that they can flexibly fuse data from different domains \cite{neverova2015moddrop, abdelaziz2017comparing}. Recently, researchers have tried to incorporate text \cite{kinoshita2015text}, bone-conducted signals \cite{yu2020time}, and visual cues \cite{koumparoulis2017exploring, wu2019time, michelsanti2019deep, iuzzolino2020av, gu2020multi, hussen2020modality} into speech applications as auxiliary and complementary information to achieve better performance. Among them, visual cues are the most common and intuitive because most devices can capture audio and visual data simultaneously. Numerous audio-visual SE (AVSE) systems have been proposed and confirmed to be effective \cite{hou2018audio, ideli2019visually, adeel2019lip, michelsanti2020overview}. In our previous work, a lite AVSE (LAVSE) approach was proposed to handle the immense visual data and potential privacy issues \cite{chuang2020lite}. The LAVSE system uses an autoencoder (AE)-based compression network along with a latent feature quantization unit \cite{wu2018training, hsu2018study} to successfully reduce the size of visual data. In practical applications, after data preprocessing, only the latent visual features extracted by the encoder of the AE are used in the processing pipeline. Since the decoder of the AE does not need to be used or disclosed, the original image is difficult to reconstruct from the visual features, and the privacy issue can be solved to a certain extent.

In this study, we intend to further explore three practical issues that are often encountered when implementing AVSE systems in real-world scenarios; they are: (1) the additional cost of processing visual data (usually much higher than the cost of processing audio data), (2) audio-visual asynchronization, and (3) low-quality visual data.

In the AVSE task, the requirement of additional visual data inevitably causes additional costs, such as computing power or memory, and visual sensors. Therefore, we need to minimize such additional costs by designing compact visual features and ensure that the system performs well under low-quality visual input. We extend the LAVSE system to an improved LAVSE (iLAVSE) system, which is formed by a multimodal convolutional RNN (CRNN) architecture in which the recurrent part is realized by implementing an LSTM layer. The audio data are provided as input directly to the SE model, while the visual input is first processed by a three-unit data compression module CRQ (C for color channel, R for resolution, and Q for bit quantization) and a pre-trained AE module. In CRQ, we adopt three data compression units: reducing the number of channels, reducing the resolution, and reducing the number of bits. The AE is formed by a deep convolutional architecture and can extract meaningful and compact representations, which are then quantized and used as input of the CRNN AVSE model. Based on the visual data compression CRQ module and AE module, the size of visual input is significantly reduced, and the privacy issue can be further addressed in iLAVSE because the original image is even more difficult to reconstruct from the visual input.

Audio-visual asynchronization is a common issue that may arise from low-quality audio-visual sensors. To handle this problem, two approaches are generally applied. One approach is to use the correlation between audio and video signals to estimate the mapping between them. For example, McAllister et al. correlated the face parameters such as mouth position to Fast Fourier Transform of the input audio signal \cite{mcallister1997lip}. In \cite{zoric2005real}, a multilayer feedforward neural network was designed to receive mel-frequency cepstral coefficients as the input and predict the viseme as the output. The other approach is to find out the time difference within the asynchronous audio-visual data. For example, based on pre-defined visual features such as bottleneck features, Marcharet et al. used a deep-neural-network-based classifier to determine a time offset \cite{marcheret2015detecting}. Chung and Zisserman proposed a two-stream structure to detect the lip-sync error and adjust the time offset \cite{chung2016out}. Halperin et al. dynamically stretched and compressed the audio signal to tackle the alignment problem \cite{halperin2019dynamic}. Rather than using DL-based model structures, we propose to handle this issue based on a data augmentation scheme.

The problem of low-quality visual data also includes the failure of the sensor to capture the visual signal. Galatas et al. evaluated the performance of audio-visual speech recognition in the presence of visual noise, such as frame drops, random Gaussian noise, and block noise \cite{galatas2011audio}. Stewart et al. evaluated the impact of MPEG-4 video compression and camera jitter on the robustness of an audio-visual speech recognition system \cite{stewart2013robust}. In this study, a practical example is the use of an AVSE system in a car-driving scenario. When the car passes through a tunnel, the visual information disappears due to the insufficient light. We solve this problem through a zero-out training scheme, which replaces the latent visual features of certain training data segments with zeros.

The proposed iLAVSE system was evaluated on the Taiwan Mandarin speech with video (TMSV) dataset\footnote{https://bio-asplab.citi.sinica.edu.tw/Opensource.html\#TMSV} \cite{chuang2020lite} and new recorded testing videos in a real-world car-driving scenario. Based on the special design of model architecture and data augmentation, iLAVSE can effectively overcome the above three issues and provide more robust SE performance than LAVSE and several related SE methods.

The remainder of this paper is organized as follows. Section \ref{sec:related} reviews related work on AVSE systems and data quantization techniques. Section \ref{sec:proposed} introduces the proposed iLAVSE system. Section \ref{sec:experiment} presents our experimental setup and results. Finally, Section \ref{sec:conclusion} provides the concluding remarks. 

\section{Related Work}
\label{sec:related}

\subsection{AVSE}

\begin{figure}[t]
	\centering
	\includegraphics[width=1.0\linewidth]{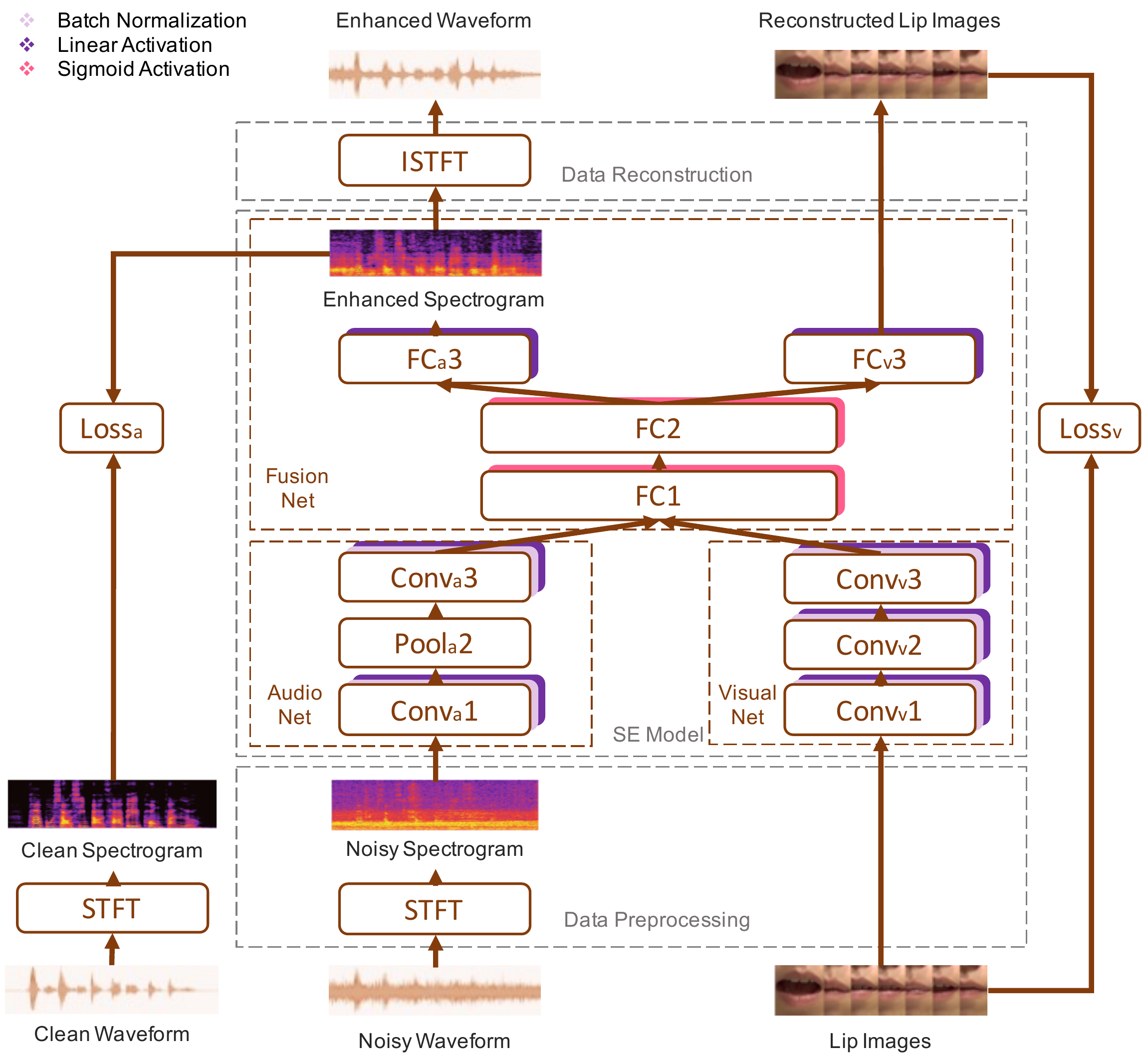}
	\caption{The AVDCNN system \cite{hou2018audio}.}
	\label{fig:AVDCNN}
\end{figure}

In this section, we review several existing AVSE systems. In \cite{hou2016audio}, a fully connected network was used to jointly process audio and visual inputs to perform SE. Since the fully connected architecture cannot effectively process visual information, the AVSE system in \cite{hou2016audio} is only slightly better than its audio-only SE counterpart. In order to further improve the performance, a multimodal deep CNN SE (termed AVDCNN) system \cite{hou2018audio} was subsequently proposed. As shown in Fig. \ref{fig:AVDCNN} (ISTFT denotes inverse short time Fourier transform; FC denotes fully connected layers; Conv denotes convolutional layers; Pool denotes max-pooling layers), the AVDCNN system consists of several convolutional layers to process audio and visual data. Experimental results show that compared with the audio-only deep CNN system, the AVDCNN system can effectively improve the SE performance. Later, Gabbay et al. proposed another AVSE model, whose architecture is similar to AVDCNN, but the visual part is not reconstructed in the output layer \cite{gabbay2018visual}. The reconstruction of the visual output in AVDCNN can guide the SE model to actually learn some useful information from the visual input, such as silence or some consonants, rather than some random information. According to our experience, the AVDCNN model with visual output performed better than the AVDCNN model without visual output. In the meantime, a looking-to-listen system was proposed, which uses estimated complex masks to reconstruct enhanced spectral features  \cite{ephrat2018looking}. In \cite{sadeghi2020audio}, a variational AE model was used as the basis model to build the AVSE system. The authors also investigated the possibility of using a strong pre-trained model for visual feature extraction and performing SE in an unsupervised manner.

Unlike audio-only SE systems, the above-mentioned AVSE systems require additional visual input, which causes additional hardware and computational costs. In addition, the use of facial or lip images may cause privacy issues. The LAVSE system \cite{chuang2020lite} has been proposed to deal with these two issues by effectively reducing the size of visual input and user identifiability. It uses an AE to extract meaningful and compact representations of visual data as the input of the SE model to reduce computational costs and appropriately solve the privacy problem in facial information. The AE in the LAVSE system is pre-trained. In \cite{chuang2020lite}, it has been shown that the AE-pre-trained framework is better than the AE-co-trained framework. In addition, the combined loss of the AE-co-trained framework consists of three losses: (1) the audio loss, (2) the visual compressed feature loss, and (3) the visual image loss. It takes time and computational cost to determine the best weights of these three losses in the AE-co-trained framework through an exhaustive search. The training process of the AE-pre-trained framework is relatively easy because there are only two losses. Moreover, in the the AE-pre-trained framework, since the AE is pre-trained in an unsupervised learning manner, it can be trained on a richer unimodal dataset.

\subsection{Data Quantization}

\begin{figure}[t]
	\centering
	\includegraphics[width=1.0\linewidth]{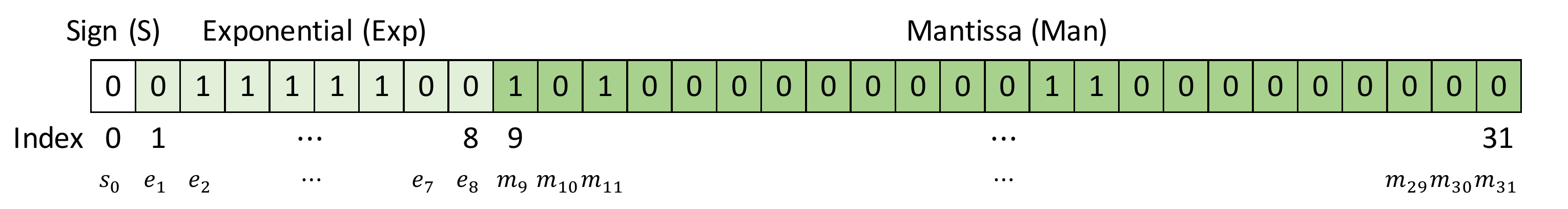}
	\caption{Single-Precision Floating-Point Format.}
	\label{fig:32fp}
\end{figure}

Quantization is a simple and effective way to reduce the size of data. Fig. \ref{fig:32fp} shows the data format of single-precision floating-point in IEEE 754 \cite{institute1985ieee}. There are 32 base-2 bits, including 1 sign bit, 8 exponential bits, and 23 mantissa bits. The decimal value of a single-precision floating-point representation is calculated as

\begin{equation}
\begin{aligned}
value_{10} &= (-1)^{S} \times 2^{(Exp_{10} - bias)} \times Man_{10}, \\
S &= s_0, \\
Exp_{2} &= e_{1}e_{2}e_{3}e_{4}e_{5}e_{6}e_{7}e_{8}, \\
Exp_{10} &= \sum_{i=1}^{8} {e_{i} \times 2^{(8-i)}}, \\
Man_{2} &= m_{9}m_{10}...m_{31}, \\
Man_{10} &= \sum_{i=9}^{31} {m_{i} \times 2^{(8-i)}},
\end{aligned}
\end{equation}

{\noindent}where the subscripts $2$ and $10$ of $value$, $Exp$, and $Man$ denote base-2 and base-10, respectively. The sign bit determines whether the value represented is positive ($S = 0$) or negative ($S = 1$). The exponential bits represent a 2's complement, which can store negative values with a bias of 127 ($2^7 - 1$). The mantissa bits are the significant figures. The decimal value of the 32-bit representation in Fig. \ref{fig:32fp} is 0.20314788.

Obviously, the representation range of values is determined by the exponential term, and the mantissa term accounts for the precision part. Therefore, quantizing the mantissa bits does not change the range, but only reduces the precision of the original value. Based on this property, an exponent-only floating-point quantized neural network (EOFP-QNN) has been proposed to reduce the mantissa bits of the SE model parameters in \cite{hsu2018study}. Experimental results have confirmed that by moderately reducing the mantissa bits, the size of the model parameters can be reduced while the overall SE capability can be improved. In this study, we followed the same idea, keeping only the sign and exponent bits, and removing all mantissa bits to perform visual data compression. 

\section{Proposed iLAVSE System}
\label{sec:proposed}

As mentioned earlier, this study investigates three practical issues: (1) the additional cost of processing visual data, (2) audio-visual data asynchronization, and (3) low-quality visual data. We propose three approaches to address these issues respectively: (1) visual data compression, (2) compensation on audio-visual asynchronization, and (3) zero-out training. By integrating the above three approaches with the CRNN AVSE architecture, the proposed iLAVSE can perform SE well even under unfavorable testing conditions. In this section, we first present the overall system of iLAVSE. Then, we describe the three issues and our solutions. 

\subsection{iLAVSE System}

\begin{figure}[t]
	\centering
	\includegraphics[width=1.0\linewidth]{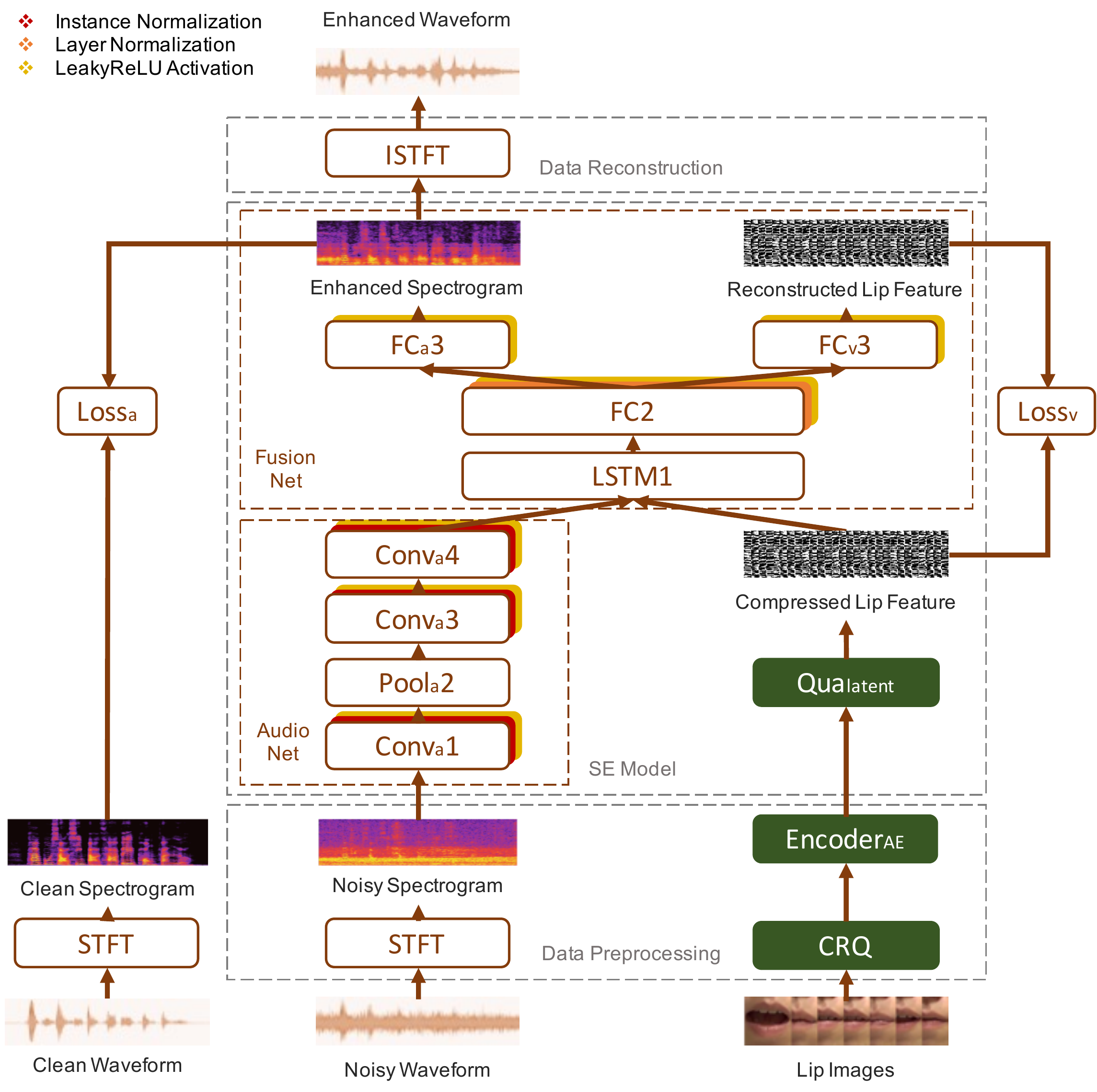}
	\caption{The proposed iLAVSE system.}
	\label{fig:whole}
\end{figure}

\begin{figure*}[t]
	\centering
	\includegraphics[width=0.8\linewidth]{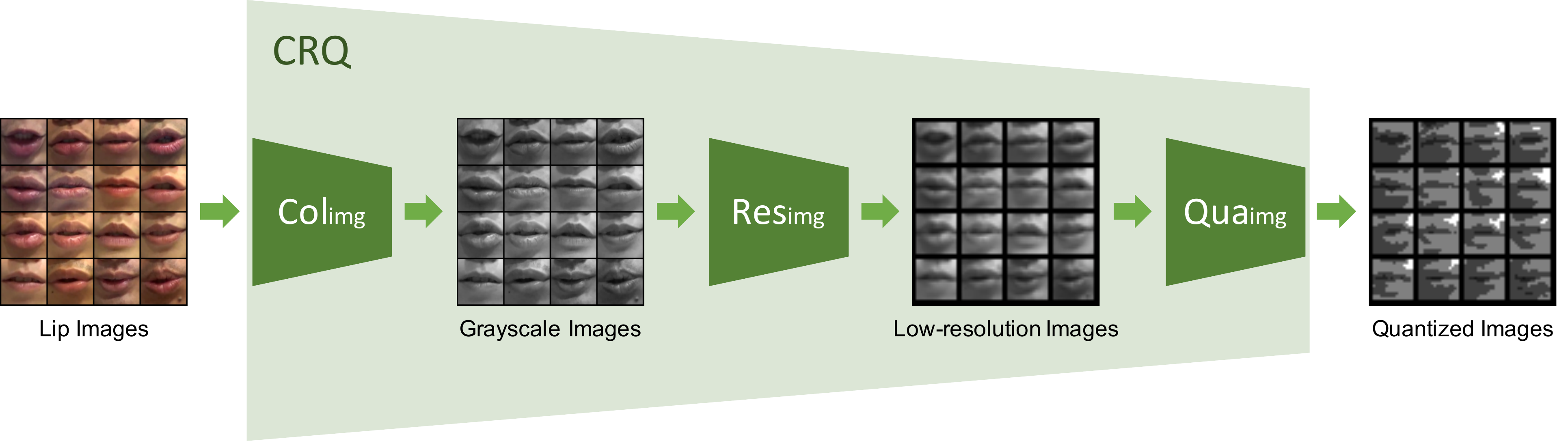}
	\caption{The proposed CRQ module.}
	\label{fig:CRQ}
\end{figure*}

The proposed iLAVSE system is demonstrated in Fig. \ref{fig:whole}. As shown in the figure, the iLAVSE system includes three stages: a data preprocessing stage, a CRNN-based AVSE stage, and a data reconstruction stage.

We have implemented three data compression functions in iLAVSE, which are outlined in green blocks in Fig. \ref{fig:whole}. CRQ is a three-unit data compression module used to compress the visual image data. As shown in Fig. \ref{fig:CRQ}, the CRQ module consists of Col{\footnotesize img}, Res{\footnotesize img}, and Qua{\footnotesize img}, denoting color channel reduction, resolution reduction, and bit quantization, respectively. Qua{\footnotesize latent} stands for the bit quantization of the latent feature extracted by Encoder{\footnotesize AE}, the encoder part of a pre-trained AE. The AE is trained by using the CRQ processed lip images as the input and the grayscale low-resolution images (cf. Fig. \ref{fig:CRQ}) of the original lip images as the output in a frame-wise manner.

In the data preprocessing stage, the waveform of the noisy data is transformed into log1p\footnote{Note that we choose the log1p feature \cite{lu2020incorporating} because its projecting range can avoid some minimum values in the data. If we take $10^{-6}$ for example and if log is applied, the projected value is $-6$; but if log1p is applied, the projected value is $0$. This characteristic enables the log1p feature to be easily normalized and trained.} spectral features ($X$) by using the short time Fourier transform (STFT), while the visual image data ($I$) are compressed and transformed into latent features ($Z$) by the CRQ module and Encoder{\footnotesize AE}.
The functions of CRQ and Encoder{\footnotesize AE} are as follows, 
\begin{equation}
\begin{aligned}
CRQ(I_{i, n}) &= Qua_{img}(Res_{img}(Col_{img}(I_{i, n}))), \\
Z_{i, n} &= Encoder_{AE}(CRQ(I_{i, n})),
\label{eq:prepro}
\end{aligned}
\end{equation}
where $i \in \{1, ..., K\}$ denotes the $i$-th training utterance, and $K$ is the number of the training utterances; $n \in \{L, ..., F-L\}$ denotes the $n$-th sample frame, $L$ is the size of the concatenated frames for a context window, and $F$ is the number of frames of the $i$-th utterance.

In the CRNN AVSE stage, the audio spectral features $X$ pass through an audio net composed of convolutional and pooling layers to extract the audio latent features ($A$), and the Qua{\footnotesize latent} unit, which will be described in Section \ref{sec:latent_feature_compression}, further quantizes the visual input $Z$ to $V$ as

\begin{equation}
\begin{aligned}
A_{i, n} &= Conv_{a}4(Conv_{a}3(Pool_{a}2(Conv_{a}1(X_{i, n-L:n+L})))), \\
V_{i, n} &= Qua_{latent}(Z_{i, n}).
\label{eq:Anet}
\end{aligned}
\end{equation}
The audio latent features $A$ and the quantized visual latent features $V$ are concatenated as $AV$, which is then sent into the fusion net and turned into $F$. Then, the fused features $F$ are decoded into the audio spectral features ($\hat{Y}$) and the visual latent features ($\hat{Z}$) respectively through a linear layer. The process is formulated as
\begin{equation}
\begin{aligned}
AV_{i, n} &= [A_{i, n}^{T}; V_{i, n-L:n+L}^{T}]^{T}, \\
F_{i, n} &= FC2(LSTM1(AV_{i, n})), \\
\hat{Y}_{i, n} &= FC_{a}3(F_{i, n}), \\
\hat{Z}_{i, n} &= FC_{v}3(F_{i, n}).
\label{eq:Fnet}
\end{aligned}
\end{equation}

During testing, the audio spectral features ($\hat{Y}$) (with the phase of the noisy speech) are reconstructed into the speech waveform using the inverse STFT in the data reconstruction stage.

%Note that we choose the log1p feature \cite{lu2020incorporating} because its projecting range can avoid some minimum values in the data. If we take $10^{-6}$ for example and if log is applied, the projected value is $-6$; but if log1p is applied, the projected value is $0$. This characteristic enables the log1p feature to be easily normalized and trained.

\subsection{Three Practical Issues and Proposed Solutions}

\subsubsection{Visual Data Compression}

For AVSE systems, the main goal is to use visual data as an auxiliary input to retrieve the clean speech signals from the distorted speech signals. However, the size of visual data is generally much larger than that of audio data, which may cause unfavorable hardware and computational costs when implementing the AVSE system. Our previous work has proven that visual data may not require very high precision, and the original image sequence can be replaced by meaningful and compact representations extracted by an AE \cite{chuang2020lite}. In this study, we further explore directly reducing the size of visual data by the CRQ compression module. The AE is directly applied to the compressed image sequence to extract a compact representation. The extracted representation is then further compressed by Qua{\footnotesize latent} and sent to the CRNN-based AVSE stage in iLAVSE. 

\paragraph{Visual Feature Extraction by a CNN-based AE}

\begin{figure*}[t]
	\centering
	\includegraphics[width=1.0\linewidth]{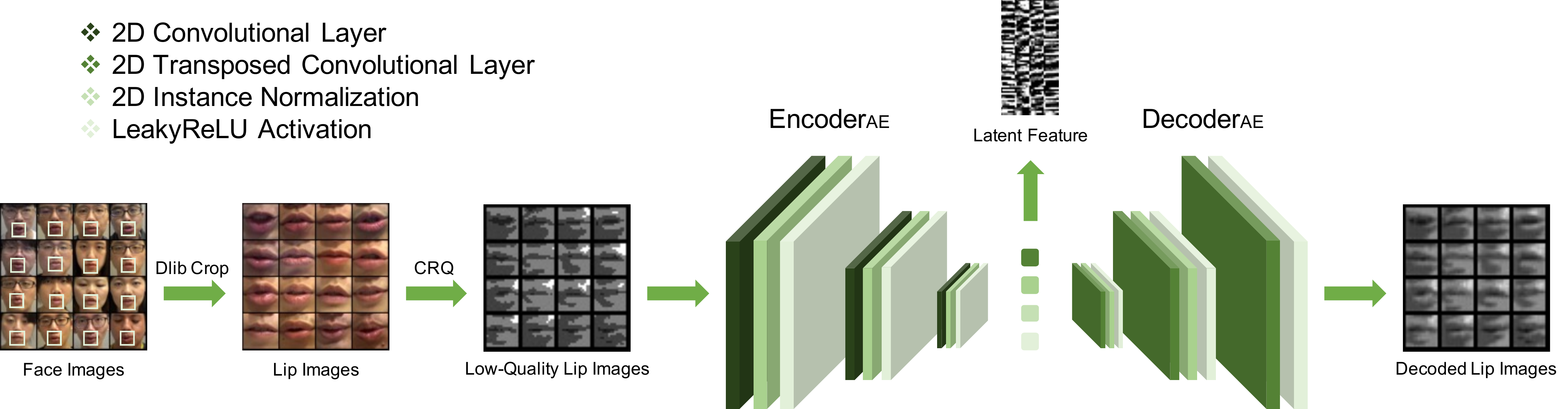}
	\caption{The AE model for visual input data compression.}
	\label{fig:AE}
\end{figure*}

As mentioned earlier, iLAVSE uses the three visual data compression units in the CRQ module, namely Col{\footnotesize img}, Res{\footnotesize img}, and Qua{\footnotesize img}, to perform color channel reduction, resolution reduction, and bit quantization, respectively. The size of the original image sequence can be notably reduced by the three units. The compressed visual data is then passed to Encoder{\footnotesize AE}, and the latent representation is used as the visual representation. As shown in Fig. \ref{fig:AE}, we use a 2D-convolution-layer-only AE to process the CRQ processed visual input. For a given CRQ processed visual input, the AE is pre-trained to reconstruct the grayscale low-resolution image (cf. Fig. \ref{fig:CRQ}) of the original lip image.

Generally, captured images are saved in RGB (three channels) or grayscale (one channel) format. Therefore, to make the iLAVSE system applicable to different scenarios, we consider both RGB and grayscale visual inputs to train the AE model. As a result, this AE model can reconstruct RGB and grayscale images. 

Furthermore, we use images with different resolutions to train the AE model. Since the lip images are about 100 to 250 pixels square, we designed three settings to reduce the resolution---\textit{64}, \textit{32}, and \textit{16} (pixels square). When using a resolution of \textit{64}, for example, the original image at sizes of 100 to 250 pixels square is resized to 64 pixels square.

For data quantization, we first quantize the values of an input image by removing the mantissa bits in the floating-point representation. To train the AE, we place the quantized and original images at the input and output, respectively. In real-world applications, the AE model can reconstruct the original visual data from the quantized version. That is to say, the color channel and size of the input and output are the same, but the number of bits is different.

\paragraph{Latent Feature Compression}
\label{sec:latent_feature_compression}

\begin{figure}[t]
	\centering
	\begin{subfigure}[t]{0.47\linewidth}
		\centering
		\includegraphics[scale=1.2]{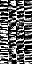}
		\caption{32-bit AE features.}
		\label{fig:latent_full}
	\end{subfigure}
	\begin{subfigure}[t]{0.47\linewidth}
		\centering
		\includegraphics[scale=1.2]{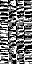}
		\caption{EOFP 3-bit AE features.}
		\label{fig:latent_eofp}
	\end{subfigure}
	\caption{Original and quantized visual latent features.}
	\label{fig:latent}
\end{figure}

\begin{figure}[t]
	\centering
	\includegraphics[width=\linewidth]{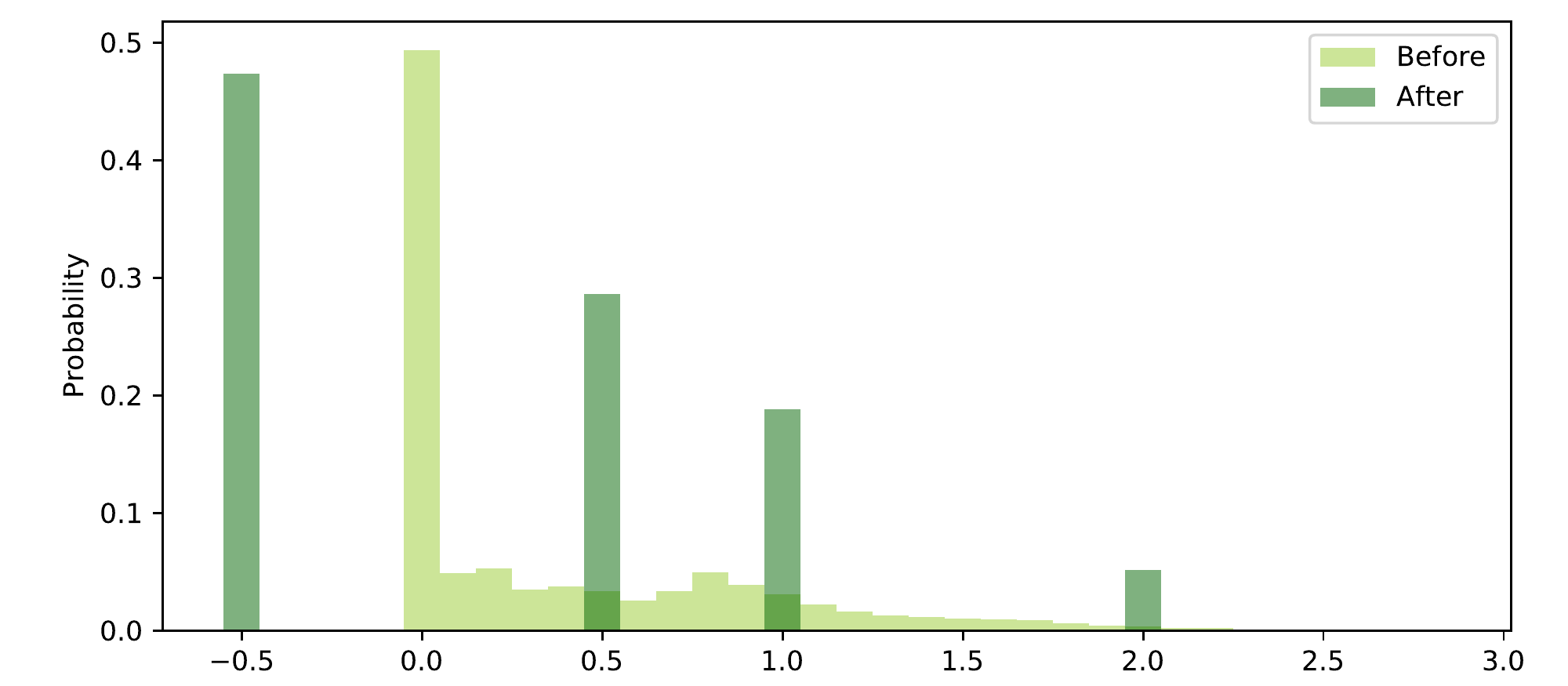}
	\caption{The distributions of visual features before and after applying Qua{\footnotesize latent}.}
	\label{fig:latent_prob}
\end{figure}

After extracting the latent feature by passing the compressed images to the AE, Qua{\footnotesize latent} in Fig. \ref{fig:whole} can further reduce the number of bits of each latent feature element. The quantized visual latent features are then used in the CRNN AVSE stage. Fig. \ref{fig:latent} shows the visual latent features before and after the Qua{\footnotesize latent} module. In real-world applications, the Encoder{\footnotesize AE} module and Qua{\footnotesize latent} unit can be installed in a low-quality visual sensor, thereby improving the online computing efficiency and greatly reducing the transmission costs. 

To further confirm that the quantized latent representation can be used to replace the original latent representation, we plotted the distributions of the latent representations before and after applying bit quantization in Fig. \ref{fig:latent_prob}. The lighter green bins represent the feature before Qua{\footnotesize latent} is applied, and the darker green bins represent the feature after Qua{\footnotesize latent} is applied. We can see that the darker green bins cover the range of the lighter green bins well, indicating that we can use the quantized latent feature to replace the original latent feature.

\subsubsection{Compensation of Audio-Visual Asynchronization}

\begin{figure}
	\centering
	\begin{subfigure}[!t]{0.49\linewidth}
		\centering
		\includegraphics[width=\linewidth]{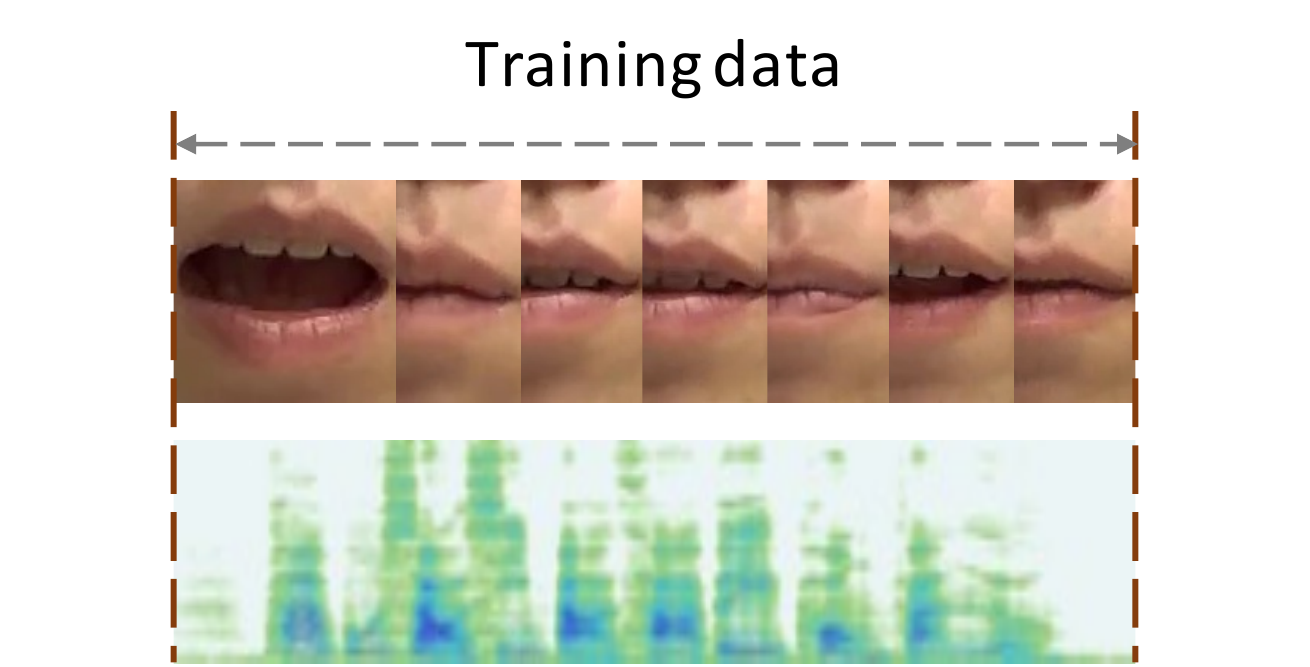}
		\caption{Synchronous.}
		\label{fig:syn}
	\end{subfigure}
	\hfill
	\begin{subfigure}[!t]{0.49\linewidth}
		\centering
		\includegraphics[width=\linewidth]{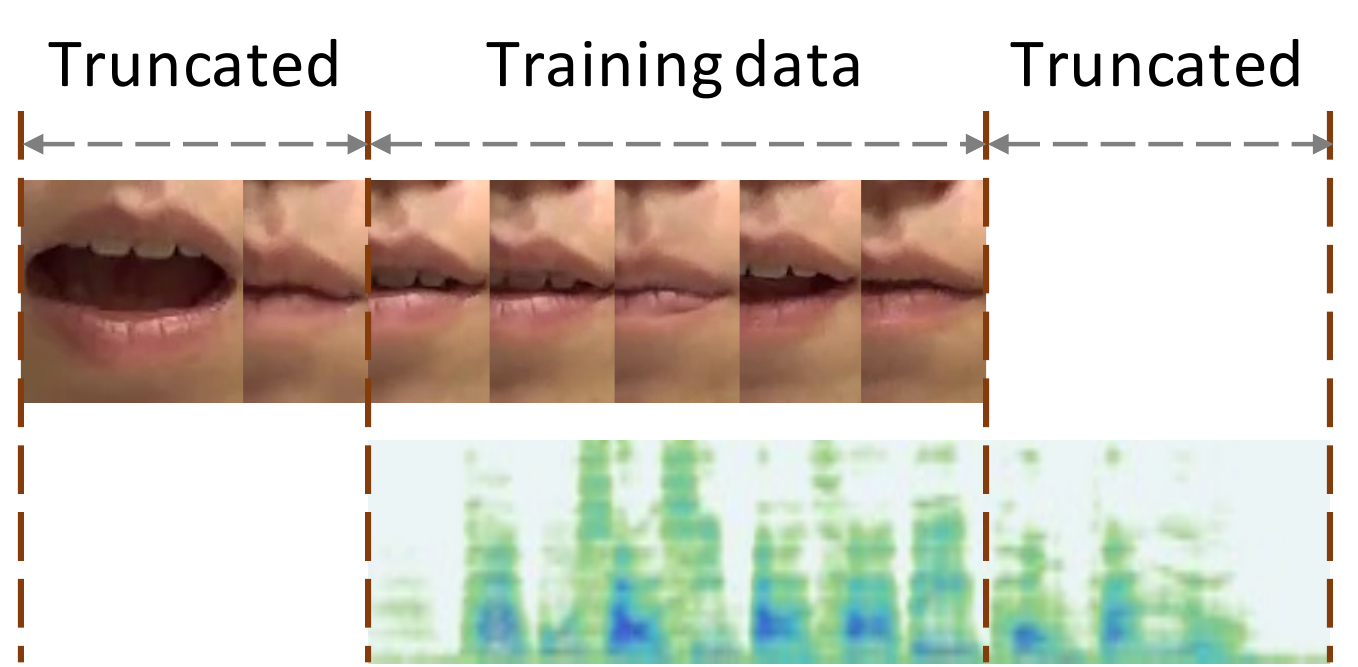}
		\caption{Asynchronous.}
		\label{fig:asyn}
	\end{subfigure}
	\caption{Synchronous and asynchronous audio and visual data.}
	\label{fig:asynchronous}
\end{figure}

Multimodal data asynchronization is a common issue in multimodal learning. We also encountered this problem when implementing the AVSE system. The ideal situation is that the audio and visual data are precisely synchronized in time. Otherwise, the auxiliary visual information may not be helpful or may even worsen the SE performance. Fig. \ref{fig:asynchronous} shows the synchronous and asynchronous situations of audio and visual data. Owing to audio-visual asynchronization, the video frames are not aligned with the speech well. In this study, we propose a data augmentation approach to alleviate this audio-visual asynchronization issue. The main idea is to artificially simulate various asynchronous audio-visual data to train the AVSE systems.

\subsubsection{Zero-Out Training}

\begin{figure}
	\centering
	\begin{subfigure}[!t]{0.47\linewidth}
		\centering
		\includegraphics[width=\linewidth]{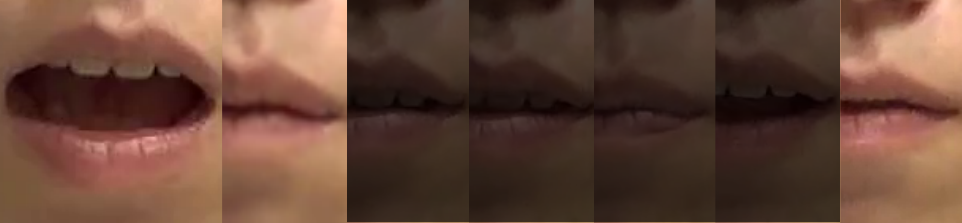}
		\caption{Low-quality lip images.}
		\label{fig:lq_image}
	\end{subfigure}
	\hfill
	\begin{subfigure}[!t]{0.47\linewidth}
		\centering
		\includegraphics[width=\linewidth]{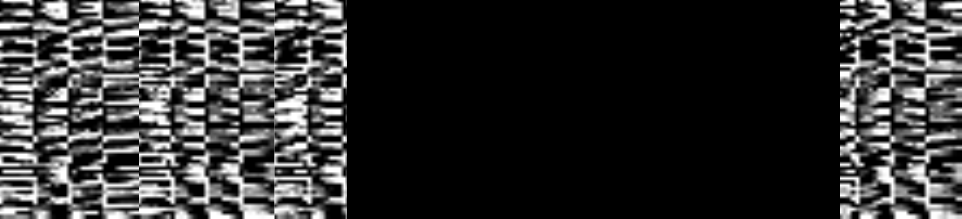}
		\caption{Low-quality latent features.}
		\label{fig:lq_latent}
	\end{subfigure}
	\caption{Low-quality visual data.}
	\label{fig:low_quality}
\end{figure}

Because visual data are regarded as an auxiliary input to the AVSE systems, a necessary requirement is that low-quality visual conditions will not degrade the SE performance. In use with poor lighting conditions, such as in a tunnel or at a night market, the quality of video frames may be poor. In Fig. \ref{fig:lq_image}, which shows an example, where a segment of frames (in the middle region) has very poor quality. Using the entire video frames directly may degrade the AVSE performance. To overcome this problem, we intend to let iLAVSE dynamically decide whether video data should be used. More specifically, when the quality of a segment of image frames is poor (which can be determined using an additional light sensor or according to the result of lip detection), iLAVSE can directly discard the visual information by replacing the visual latent features of low-quality frames with zeros, as shown in Fig. \ref{fig:lq_latent}. In order to make iLAVSE have the ability to process audio information alone, in the training phase, we prepare training data by replacing the visual latent features of the visual frames of certain segments with zeros. In this way, when the video quality is low, iLAVSE can perform SE based on audio input only, without considering visual information.

Note that this study only considers low-quality situations that occur in consecutive frame segments, not in sporadic frames; this situation is common in car-driving scenarios. We believe that the proposed zero-out training method is suitable for other low-quality visual data scenarios, because it is a common idea to set the visual input to zeros when the video quality is poor. In the future, we will conduct experiments to verify this idea in other real-world scenarios. In addition, the focus of this study is to verify whether the proposed iLAVSE system can function well even when some visual data are discarded. The criterion that can best determine whether visual information should be discarded will be our future work.

\section{Experiments}
\label{sec:experiment}

This section presents the experimental setup and results. Two standardized evaluation metrics were used to evaluate the SE performance: perceptual evaluation of speech quality (PESQ) \cite{rix2001perceptual} and short-time objective intelligibility measure (STOI) \cite{taal2011algorithm}. PESQ was developed to evaluate the quality of processed speech. The score ranges from -0.5 to 4.5. A higher PESQ score indicates that the enhanced speech has better speech quality. STOI was designed to evaluate the speech intelligibility. The score typically ranges from 0 to 1. A higher STOI value indicates better speech intelligibility. 

\begin{figure}
	\centering
	\begin{subfigure}[!t]{0.6\linewidth}
		\centering
		\includegraphics[width=\linewidth]{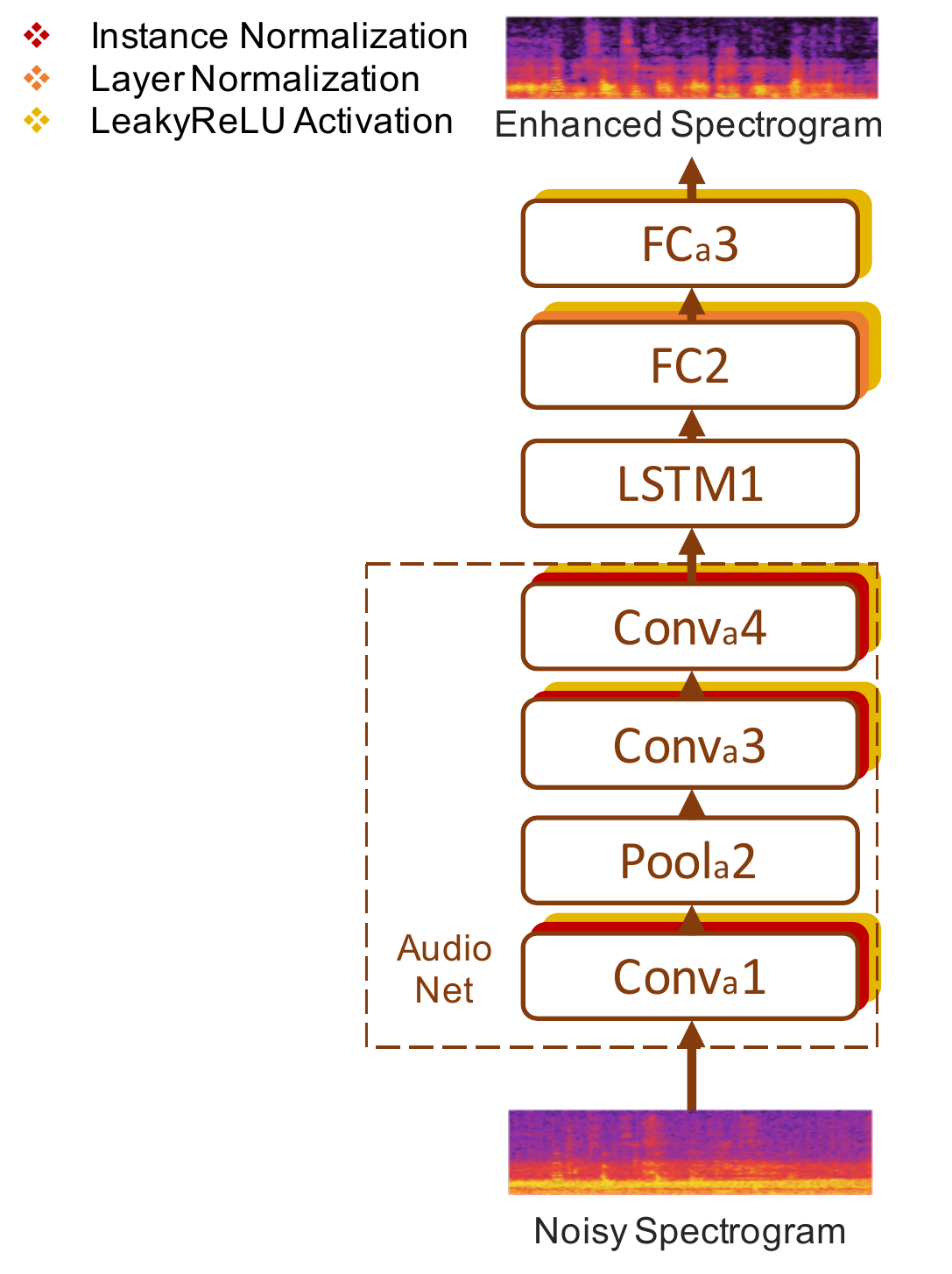}
		\caption{Audio-only SE.}
		\label{fig:ao}
	\end{subfigure}
	\par\medskip
	\begin{subfigure}[!t]{0.8\linewidth}
		\centering
		\includegraphics[width=\linewidth]{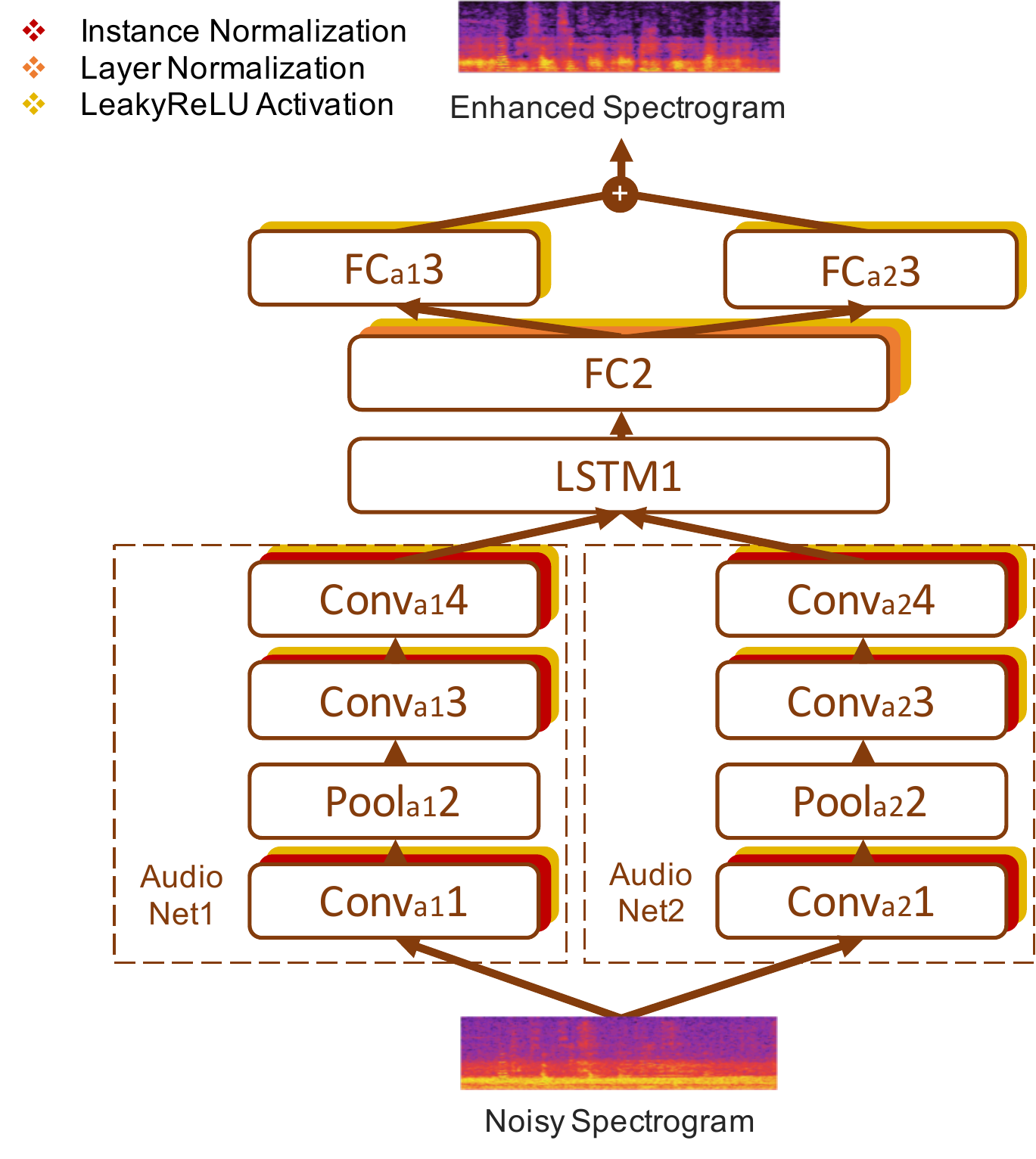}
		\caption{Dual-path-audio-only SE.}
		\label{fig:aod}
	\end{subfigure}
	\caption{Architectures of two audio-only SE systems.}
	\label{fig:baseline}
\end{figure}

Two audio-only baseline SE systems were implemented for comparison. Their model architectures are illustrated in Fig. \ref{fig:baseline}. Fig. \ref{fig:ao} is a system with the visual part in the iLAVSE system deleted, and Fig. \ref{fig:aod} is a system with a dual-path audio model. The additional audio net in Fig. \ref{fig:aod} is to increase the number of model parameters to be the same as in the iLAVSE model. This system tests whether additional improvements can be achieved by simply increasing the number of model parameters. 

The loss function for training iLAVSE is based on the mean square error computed from both the audio and visual parts,

\begin{equation}
\begin{aligned}
Loss_{a} &= \frac{1}{KF}\sum_{i=1}^{K}\sum_{n=1}^{F} {\lvert\lvert \hat{Y}_{i, n} - Y_{i, n} \rvert\rvert}^2, \\
Loss_{v} &= \frac{1}{KF}\sum_{i=1}^{K}\sum_{n=1}^{F} {\lvert\lvert \hat{Z}_{i, n} - Z_{i, n} \rvert\rvert}^2, \\
Loss &= Loss_{a} + \mu \times Loss_{v},
\end{aligned}
\end{equation}

{\noindent}where $\mu$ is empirically determined as $10^{-3}$. For training the two audio-only SE systems, $Loss_a$ is used.

In this study, all the SE models were implemented using the PyTorch \cite{paszke2019pytorch} library. The optimizer is Adam \cite{kingma2015adam} with a learning rate of $5 \times 10^{-5}$. The training batch size was set to 32.

\subsection{Experimental Setup}

In this section, the details of the dataset and the implementation steps of iLAVSE and other SE systems are introduced. 

\subsubsection{Dataset}

We evaluated the proposed system on the TMSV dataset\footnote{https://bio-asplab.citi.sinica.edu.tw/Opensource.html\#TMSV}. The dataset contains video recordings of 18 native speakers (13 males and 5 females), each speaking 320 utterances of Mandarin sentences, with the script of the Taiwan Mandarin hearing in noise test \cite{huang2005development}. Each sentence has 10 Chinese characters, and the length of each utterance is approximately 2–4 seconds. The utterances were recorded in a recording studio with sufficient light, and the speakers were filmed from the front view. The video was recorded at a resolution of 1920 pixels $\times$ 1080 pixels at 50 frames per second. The audio was recorded at a sampling rate of 48 kHz.

In this study, considering gender balance, we decided not to use all 18 speakers from TMSV. We selected the video files from 8 speakers (4 males and 4 females) to form the training set. For each speaker, among the 320 utterances, the 1-st to the 200-th utterances were selected. The utterances were artificially corrupted by 100 types of noise \cite{hu2004100} at 5 different signal-to-noise ratio (SNR) levels, from -12 dB to 12 dB with a step of 6 dB. This process yielded about 600 hours of noisy utterances. Considering that 600 hours of training data would take too much training time, we randomly sampled 12,000 noisy utterances as a 9-hour training set. The 201-st to 320-th video recordings of 2 other speakers (1 male and 1 female) were used to form the testing set. Six types of noise were selected, which are common in car-driving scenarios, including baby cry, engine noise, background talkers, music, pink noise, and street noise. We artificially generated noisy utterances by contaminating the clean testing speech with these 6 types of noise at 4 low SNR levels, including -1, -4, -7, and -10 dB, which are around the SNR levels mentioned in \cite{park2014voice}. This process produced 5,760 testing noisy utterances for a total of about 4 hours. The speakers, speech contents, noise types, and SNR levels were all mismatched in the training and testing sets. 

\subsubsection{Audio and Visual Feature Extraction}

The recorded speech signals were downsampled to 16 kHz and mixed into monaural waveforms. The speech waveforms were converted into spectrograms with STFT. The window size of STFT was 512, corresponding to 32 milliseconds. The hop length was 320, so the interval between each frame was 20 milliseconds. The audio data was formatted at 50 frames per second and was aligned with the video data. For each speech frame, the log1p magnitude spectrum \cite{lu2020incorporating} was extracted, and the value was normalized to zero mean and unit standard deviation. The normalization process was conducted at the utterance level; that is, the mean and standard deviation vectors were calculated on all frames of an utterance. The length of the context window was 5, i.e., $\pm$2 frames were concatenated to the central frame. Accordingly, the dimension of the final frame-based audio feature vector was 257 $\times$ 5.

For each frame in the video, the contour of the lips was detected using a 68-point facial landmark detector with Dlib \cite{dlib09}, and the RGB channels were retained. The extracted lip images were approximately 100 pixels square to 250 pixels square. The AE was trained on the lip images in the training set. The latent representation (2048-dimensional) of AE were used as the visual input to the CRNN-based AVSE stage. Same as the audio feature, $\pm$2 frames were concatenated to the central frame. Therefore, the dimension of the frame-based visual feature vector was 2048 $\times$ 5.

\subsection{Experimental Result}

\subsubsection{AVSE Versus Audio-Only SE}

\begin{table}[!b]
	\begin{center}
		\begin{tabular}{c|cc}
			\hline
			\hline
			                                  & \textbf{PESQ}  & \textbf{STOI}  \\
			\hline
			\textbf{Noisy}                    & 1.001          & 0.587          \\
			\textbf{AOSE}                     & 1.282          & 0.616          \\
			\textbf{AOSE(DP)}                 & 1.283          & 0.610          \\
			\textbf{AVDCNN}                   & 1.337          & 0.641          \\
			\textbf{LAVSE(AE)}                & 1.374          & \textbf{0.646} \\
			\textbf{LAVSE(AE+EOFP4bits)}      & 1.358          & 0.643          \\
			\textbf{iLAVSE(CRQ)}              & 1.387          & 0.639          \\
			\textbf{iLAVSE(CRQ+AE)}           & 1.398          & 0.641          \\
			\textbf{iLAVSE(CRQ+AE+EOFP3bits)} & \textbf{1.410} & 0.641          \\
			\hline
			\hline
		\end{tabular}
		\caption{Average PESQ and STOI scores of the two audio-only SE systems and the AVSE systems over SNRs of -1, -4, -7 and -10 dB.}
		\label{tab:baseline}
	\end{center}
\end{table}

The two audio-only SE systems shown in Fig. \ref{fig:baseline} were used as the baselines. The results of the audio-only SE (denoted as AOSE) and dual-path audio-only SE (denoted as AOSE(DP)) systems are shown in Table \ref{tab:baseline}. As mentioned earlier, AOSE(DP) has a similar number of model parameters to LAVSE. From the results in Table \ref{tab:baseline}, we note that AOSE and AOSE(DP) yield similar performance in terms of PESQ and STOI. The result suggests that the additional path with extra parameters cannot provide improvements for the audio-only SE system in this task. Table \ref{tab:baseline} also lists the results of the proposed iLAVSE and two existing AVSE systems, namely AVDCNN \cite{hou2018audio} and LAVSE \cite{chuang2020lite}. LAVSE(AE) denotes the LAVSE system with AE, while LAVSE(AE+EOFP4bits) denotes the LAVSE system with both AE and the latent feature quantization unit Qua{\footnotesize latent} for 4 bits of EOFP. The proposed CRQ module can also be regarded as a coding method that can reduce user identifiability in the image domain. The iLAVSE system with CRQ but without AE is denoted as iLAVSE(CRQ), while the iLAVSE system with CRQ and AE is denoted as iLAVSE(CRQ+AE). In addition, iLAVSE(CRQ+AE+EOFP3bits) stands for the system including CRQ, AE, and Qua{\footnotesize latent} for 3 bits of EOFP. The results show that the systems with compression modules of CRQ and AE and the quantization unit Qua{\footnotesize latent} can maintain SE performance comparable to LAVSE(AE). Compared to AOSE and AOSE(DP), all the AVSE systems yield higher PESQ and STOI scores, confirming the effectiveness of incorporating visual data into the SE system.

\subsubsection{Visual Data Compression}

In this set of experiments, we examined the ability of iLAVSE to incorporate compressed visual data. As shown in Fig. \ref{fig:whole}, the visual data preprocessing is carried out by a CRQ module, which implements three units: Col{\footnotesize img}, Res{\footnotesize img}, and Qua{\footnotesize img}. Then, after the latent representation is extracted by Encoder{\footnotesize AE}, Qua{\footnotesize latent} further quantizes the bits of the latent representation. In other words, there are four units that perform visual data reduction. We represent the entire reduction process as \{Col{\footnotesize img}, Res{\footnotesize img}, Qua{\footnotesize img}, Qua{\footnotesize latent}\} = \{A, B, C, D\}, where A is either \textit{RGB} or \textit{GRAY} (for grayscale), B denotes the image resolution, C indicates the image data quantization, and D stands for the latent feature quantization.

\begin{table}[!b]
	\begin{center}
		\begin{tabular}{c|cc|cc}
			\hline
			\hline
			\multicolumn{1}{c|}{}                  & \multicolumn{2}{c|}{\textbf{PESQ}}     & \multicolumn{2}{c}{\textbf{STOI}}    \\
			                                       & \textbf{R}            & \textbf{G}     & \textbf{R}          & \textbf{G}     \\
			\hline
			\multicolumn{1}{c|}{\textbf{AOSE(DP)}} & \multicolumn{2}{c|}{1.283}             & \multicolumn{2}{c}{0.610}            \\
			\textbf{iLAVSE \textit{64}}            & \underline{1.374}     & \textbf{1.378} & \underline{0.646}   & 0.646          \\
		    \textbf{iLAVSE \textit{32}}            & 1.371                 & 1.375          & 0.644               & 0.645          \\
			\textbf{iLAVSE \textit{16}}            & 1.374                 & 1.358          & 0.646               & \textbf{0.649} \\
			\hline
			\hline
		\end{tabular}
		\caption{The performance of iLAVSE using lip images with reduced channel numbers and resolutions, \textbf{R}: \{\textit{RGB}\} and \textbf{G}: \{\textit{GRAY}\}. The underlined scores are the same as those of LAVSE in Table \ref{tab:baseline} because the iLAVSE with the \{\textit{RGB}, \textit{64}\} setup is equivalent to LAVSE.}
		\label{tab:AE_output}
	\end{center}
\end{table}

We evaluated iLAVSE with different types of compressed visual data. The results are listed in Table \ref{tab:AE_output}. From the table, we first see that iLAVSE outperforms AOSE(DP) in terms of PESQ and STOI with different compressed visual data. Moreover, compared to LAVSE (the underlined scores), we note that iLAVSE can still achieve comparable performance even though the resolution of the visual data has been notably reduced. For example, the \{\textit{GRAY}, \textit{16}\} case in Table \ref{tab:AE_output} strikes a good balance between the data compression ratio of 48 (($3 \div 1) \times ((64 \times 64) \div (16 \times 16)$)) and the PESQ and STOI scores. Therefore, we decided to use \{\textit{GRAY}, \textit{16}\} as a representative setup in the following discussion.

\begin{figure}
	\centering
	\begin{subfigure}[!t]{0.47\linewidth}
		\centering
		\includegraphics[scale=1.]{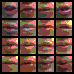}
		\caption{\{\textit{RGB}, \textit{16}, 5bits(i)\} input.}
	\end{subfigure}
	\begin{subfigure}[!t]{0.47\linewidth}
		\centering
		\includegraphics[scale=1.]{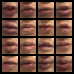}
		\caption{\{\textit{RGB}, \textit{16}, 5bits(i)\} output.}
	\end{subfigure}
	\par\medskip
	\begin{subfigure}[!t]{0.47\linewidth}
		\centering
		\includegraphics[scale=1.]{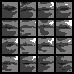}
		\caption{\{\textit{GRAY}, \textit{16}, 5bits(i)\} input.}
	\end{subfigure}
	\begin{subfigure}[!t]{0.47\linewidth}
		\centering
		\includegraphics[scale=1.]{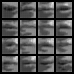}
		\caption{\{\textit{GRAY}, \textit{16}, 5bits(i)\} output.}
	\end{subfigure}
	\caption{AE lip images in 5 bits (1 sign bit and 4 exponential bits).}
	\label{fig:lip_5bits}
\end{figure}

\begin{table}[!b]
	\begin{center}
		\begin{tabular}{c|cc|cc}
			\hline
			\hline
			\multicolumn{1}{c|}{} & \multicolumn{2}{c|}{\textbf{PESQ}} & \multicolumn{2}{c}{\textbf{STOI}}  \\
			\textbf{Total bits}   & \textbf{R}        & \textbf{G}     & \textbf{R}        & \textbf{G}     \\
			\hline
			\textbf{1}            & 1.333             & 1.296          & 0.619             & 0.615          \\
			\textbf{3}            & 1.250             & 1.295          & 0.628             & 0.613          \\
			\textbf{5}            & 1.361             & \textbf{1.398} & 0.644             & 0.641          \\
			\textbf{7}            & 1.374             & 1.379          & 0.640             & 0.644          \\
			\textbf{9}            & 1.386             & 1.387          & 0.642             & 0.642          \\
			\textbf{32}           & \underline{1.374} & 1.358          & \underline{0.646} & \textbf{0.649} \\
			\hline
			\hline
		\end{tabular}
		\caption{The performance of iLAVSE with or without image quantization (the original image is with 32 bits), \textbf{R}: \{\textit{RGB}, \textit{64}\} and \textbf{G}: \{\textit{GRAY}, \textit{16}\}. The underlined scores are the same as those of LAVSE in Table \ref{tab:baseline}.}
		\label{tab:qi_score}
	\end{center}
\end{table}

Next, we investigated quantized images. The input and output (reconstructed) images in \textit{RGB} and \textit{GRAY} are shown in the left and right columns in Fig. \ref{fig:lip_5bits}, respectively. The original 32-bit images were reduced to 5-bit images (1 sign bit and 4 exponential bits). From the figures, we observe that the AE can reconstruct the quantized image well. We also evaluated iLAVSE with the quantized images. The results are shown in Table \ref{tab:qi_score}. The PESQ and STOI scores reveal that when the numerical precision of the input image is reduced to 5 bits (1 sign bit and 4 exponential bits), iLAVSE still maintains satisfactory performance. When the number of bits is  further reduced, the PESQ and STOI scores both decrease notably. Compared to LAVSE that uses raw visual data, the overall compression ratio $R_{comp}$ of the CRQ module from \{\textit{RGB}, \textit{64}, 32bits(i)\} to \{\textit{GRAY}, \textit{16}, 5bits(i)\} is 307.2 times, which is calculated as follows,

\begin{equation}
\begin{aligned}
R_{comp} &= R_{color} \times R_{res} \times R_{Qua}, \\
R_{color} &= \frac{3}{1}, \\
R_{res} &= \frac{64^2}{16^2}, \\
R_{Qua} &= \frac{32}{5}, \\
R_{comp} &= \frac{3}{1} \times \frac{64^2}{16^2} \times \frac{32}{5} = 307.2.
\end{aligned}
\end{equation}

\subsubsection{Latent Feature Quantization}

In this set of experiments, we investigated the impact of the bit quantization in the Qua{\footnotesize latent} unit on the visual latent representation. We intended to use fewer bits to represent the original 32-bit latent representation. The compressed representation was used as the visual feature input of the AVSE model. In Fig. \ref{fig:latent_full} and Fig. \ref{fig:latent_eofp}, the latent representations of lip features before and after applying data quantization (from 32 bits to 3 bits) are depicted. As can be seen from the figures, the speaker identity cannot be fully recovered from the encoded features. Since the original images cannot be reconstructed from the compact latent features without the matched decoder and inverse EOFP procedure, the user’s privacy can be protected in the AVSE stage, thereby moderately addressing the privacy problem.

\begin{table}[!b]
	\begin{center}
		\begin{tabular}{c|cc|cc}
			\hline
			\hline
			\multicolumn{1}{c|}{} & \multicolumn{2}{c|}{\textbf{PESQ}} & \multicolumn{2}{c}{\textbf{STOI}} \\
			\textbf{Total bits}   & \textbf{R} & \textbf{G}            & \textbf{R}     & \textbf{G}       \\
			\hline
			\textbf{1}            & 1.365      & 1.374                 & 0.642          & 0.642            \\
			\textbf{3}            & 1.337      & 1.410                 & 0.642          & 0.641            \\
			\textbf{5}            & 1.343      & \textbf{1.413}        & 0.643          & 0.641            \\
			\textbf{7}            & 1.357      & 1.391                 & 0.643          & 0.641            \\
			\textbf{9}            & 1.362      & 1.373                 & 0.643          & 0.643            \\
			\textbf{32}           & 1.374      & 1.398                 & \textbf{0.646} & 0.641            \\
			\hline
			\hline
		\end{tabular}
		\caption{The performance of iLAVSE with or without latent quantization, \textbf{R}: \{\textit{RGB}, \textit{64}, 32bits(i)\} and \textbf{G}: \{\textit{GRAY}, \textit{16}, 5bits(i)\} (1 sign bit + 4 exponential bits).}
		\label{tab:double_performance}
	\end{center}
\end{table}

We further evaluated iLAVSE with latent representation quantization. The number of bits was reduced from 32 to 1, 3, 5, 7 and 9 (1 sign bit and 0, 2, 4, 6, and 8 exponential bits). The results are listed in Table \ref{tab:double_performance}. From the table, we can note that for different types of visual input, latent representations with different levels of quantization provide similar performance in terms of PESQ and STOI. For example, when quantizing the latent representation to 3 bits, PESQ = 1.410 and STOI = 0.641 under the condition of \{\textit{GRAY}, \textit{16}, 5bits(i)\}, which are much better than the performance of AOSE(DP) (PESQ = 1.283 and STOI = 0.610) and comparable to the performance of LAVSE (PESQ = 1.374 and STOI = 0.646).

\subsubsection{Further Analysis}

\begin{figure}
	\centering
	\begin{subfigure}[!t]{1.0\linewidth}
		\centering
		\includegraphics[width=\linewidth]{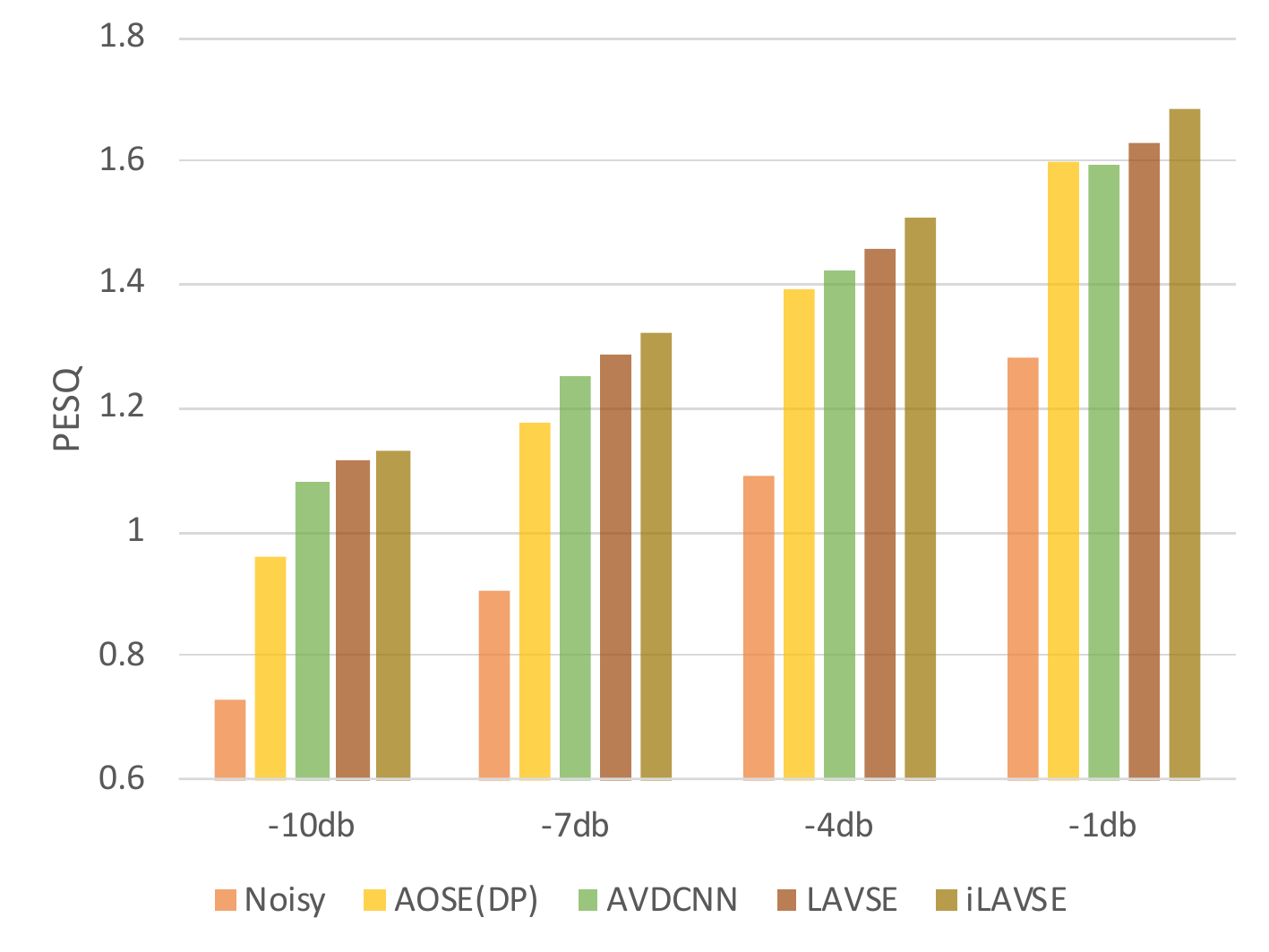}
		\caption{PESQ.}
	\end{subfigure}
	\par\medskip
	\begin{subfigure}[!t]{1.0\linewidth}
		\centering
		\includegraphics[width=\linewidth]{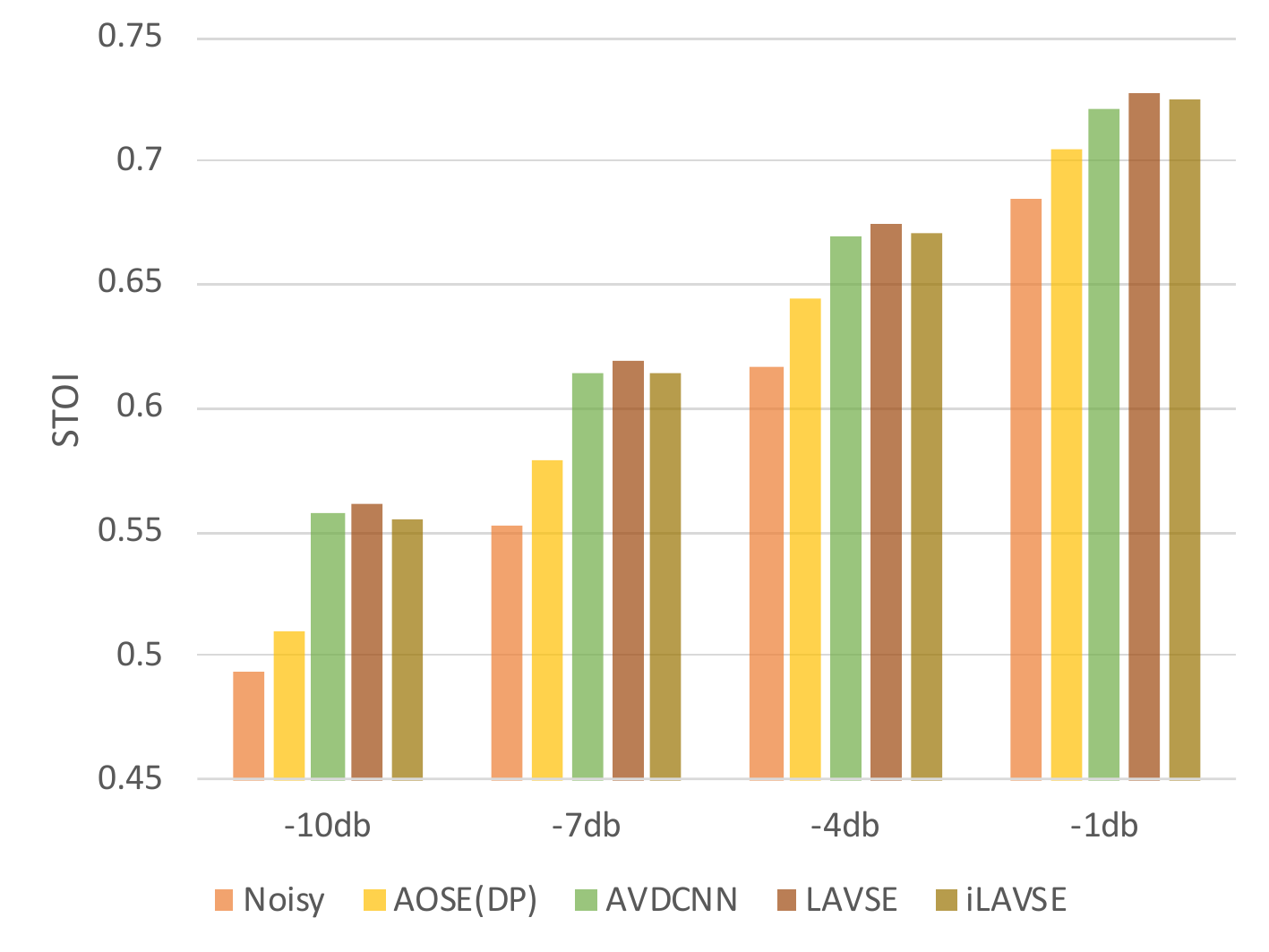}
		\caption{STOI.}
	\end{subfigure}
	\caption{The performance of different SE systems at different SNR levels. LAVSE: \{\textit{RGB}, \textit{64}, 32bits(i), 32bits(l)\}, iLAVSE: \{\textit{GRAY}, \textit{16}, 5bits(i), 3bits(l)\}.}
	\label{fig:snr_performance}
\end{figure}

In this set of experiments, we evaluated the SE systems compared in this study with different SNR levels. For AVDCNN, we used the original high-quality images as visual input. For LAVSE, we used the \{\textit{RGB}, \textit{64}, 32bits(i), 32bits(l)\} setup. For iLAVSE, we used \{\textit{GRAY}, \textit{16}, 5bits(i), 3bits(l)\}, where (i) and (l) denote the quantization unit applied to the images and the latent features, respectively. The PESQ and STOI scores for different SNR levels are shown in Fig. \ref{fig:snr_performance}, where the x-axis represents the SNR level. It can be seen from the figure that all four SE systems have higher PESQ and STOI scores than the ``Noisy'' speech. In addition, the iLAVSE system is always better than the other three SE systems at different SNR levels in terms of PESQ, and maintains satisfactory performance in terms of STOI. Through the results of -1 dB and -10 dB, we can see that visual information becomes more useful for SE tasks when the SNR decreases.

\begin{figure}
	\centering
	\begin{subfigure}[!t]{1.0\linewidth}
		\centering
		\includegraphics[width=\linewidth]{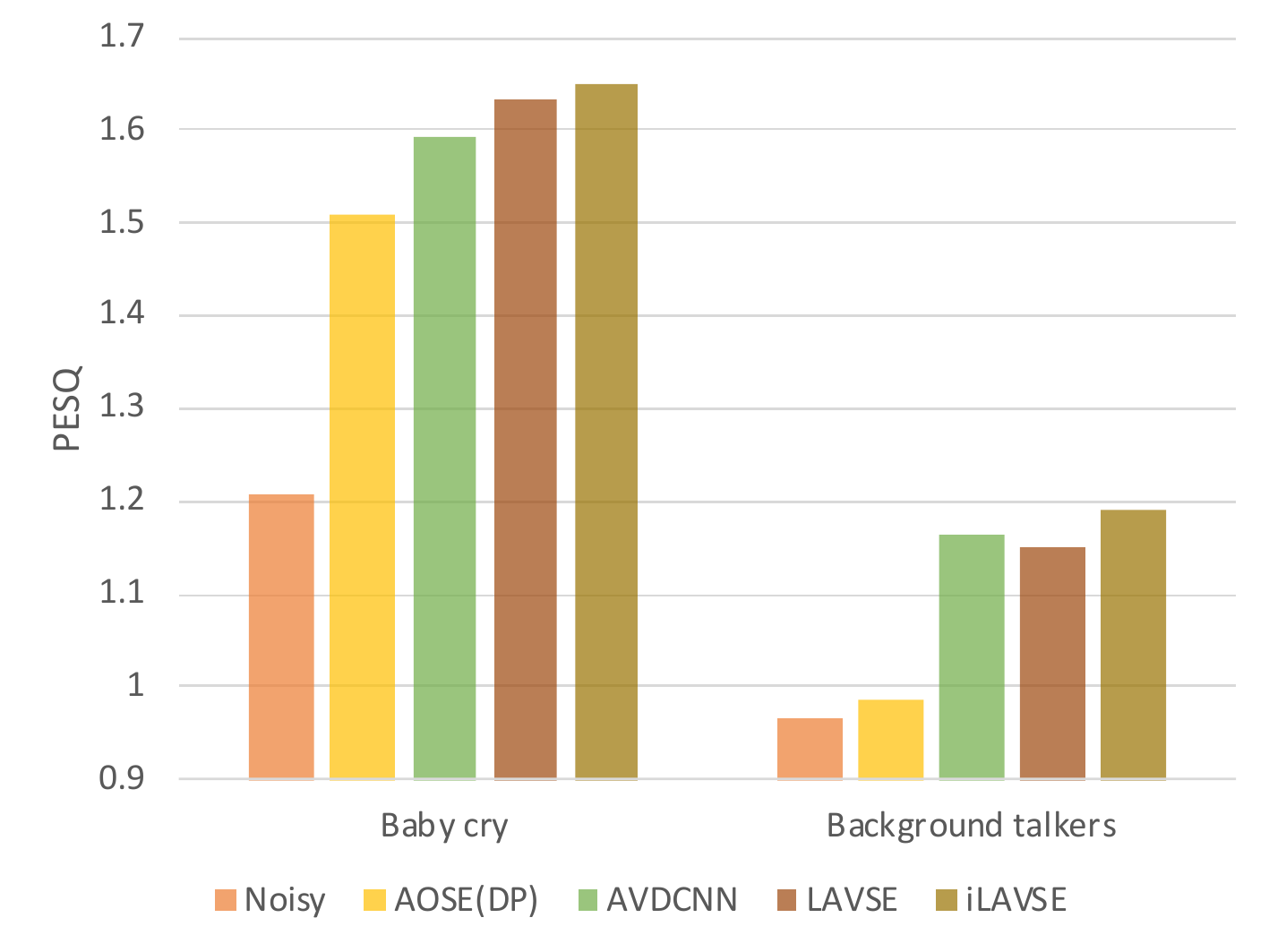}
		\caption{PESQ.}
	\end{subfigure}
	\par\medskip
	\begin{subfigure}[!t]{1.0\linewidth}
		\centering
		\includegraphics[width=\linewidth]{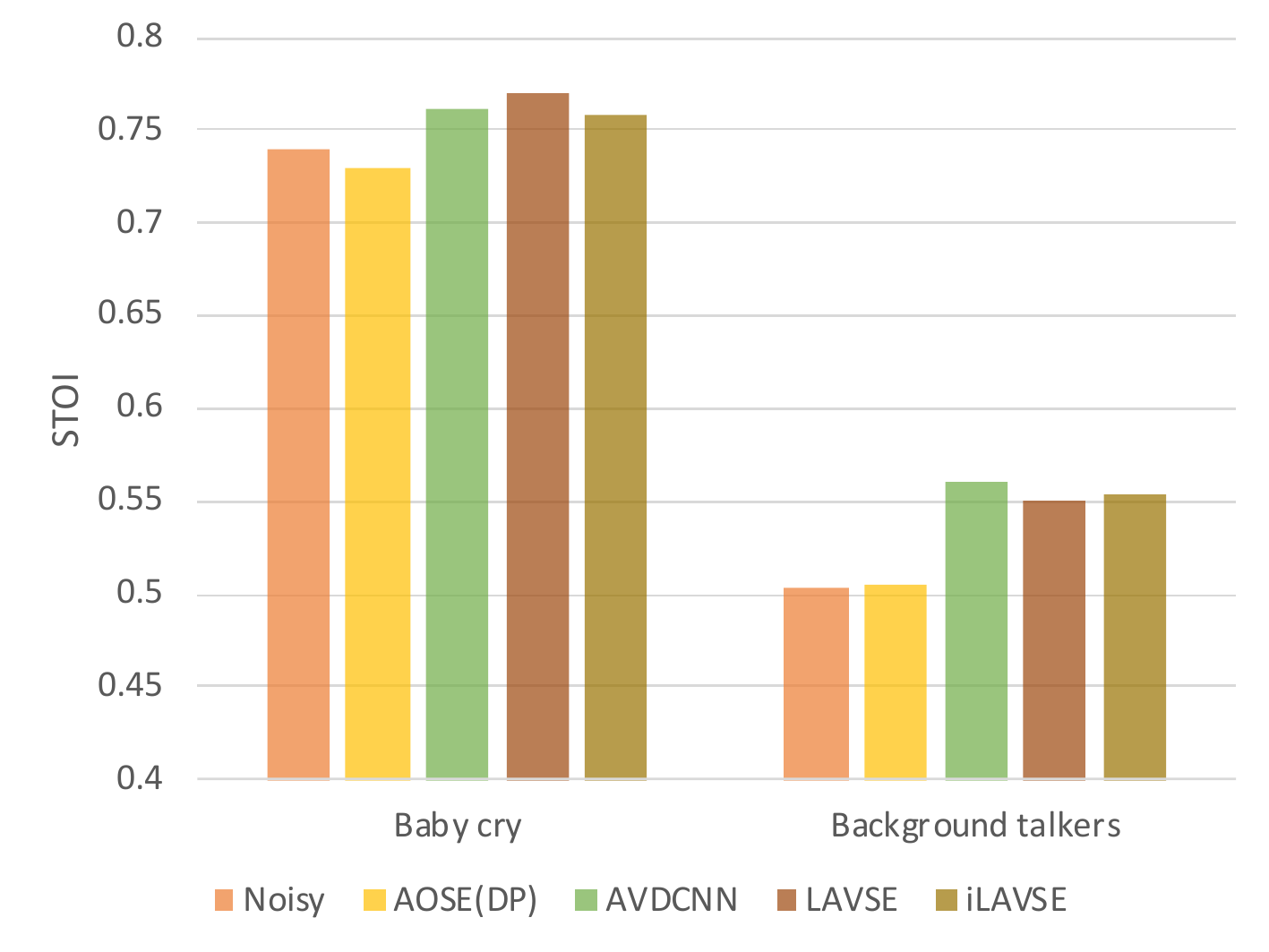}
		\caption{STOI.}
	\end{subfigure}
	\caption{The performance of different SE systems on different human-voiced noises. LAVSE: \{\textit{RGB}, \textit{64}, 32bits(i), 32bits(l)\}, iLAVSE: \{\textit{GRAY}, \textit{16}, 5bits(i), 3bits(l)\}.}
	\label{fig:noise_performance}
\end{figure}

\begin{table}[!b]
    \centering
    \begin{subtable}[!b]{0.45\textwidth}
        \centering
        \begin{tabular}{c|cc|cc}
			\hline
			\hline
			\multicolumn{1}{c|}{} & \multicolumn{2}{c|}{\textbf{PESQ}}  & \multicolumn{2}{c}{\textbf{STOI}}   \\
			\textbf{SNRs}         & \textbf{AOSE(DP)} & \textbf{iLAVSE} & \textbf{AOSE(DP)} & \textbf{iLAVSE} \\
			\hline
			\textbf{Poor}         & 1.387             & 1.544           & 0.699             & 0.734           \\
			\textbf{Low}          & 1.629             & 1.757           & 0.760             & 0.783           \\
			\textbf{Mild}         & 1.886             & 1.966           & 0.812             & 0.823           \\
			\hline
			\hline
		\end{tabular}
		\caption{Baby cry.}
	\end{subtable}
	\par\medskip
    \begin{subtable}[!b]{0.45\textwidth}
        \centering
        \begin{tabular}{c|cc|cc}
			\hline
			\hline
			\multicolumn{1}{c|}{} & \multicolumn{2}{c|}{\textbf{PESQ}}  & \multicolumn{2}{c}{\textbf{STOI}}   \\
			\textbf{SNRs}         & \textbf{AOSE(DP)} & \textbf{iLAVSE} & \textbf{AOSE(DP)} & \textbf{iLAVSE} \\
			\hline
			\textbf{Poor}         & 0.793             & 1.009           & 0.435             & 0.487           \\
			\textbf{Low}          & 1.183             & 1.372           & 0.575             & 0.621           \\
			\textbf{Mild}         & 1.575             & 1.733           & 0.702             & 0.733           \\
			\hline
			\hline
		\end{tabular}
		\caption{Background talkers.}
	\end{subtable}
	\hfill
	\caption{The performance of AOSE(DP) and iLAVSE on different human-voiced noises at different SNR levels. Poor: \mbox{-10db} and -7db, Low: -4 and -1db, Mild: 2db and 5db. iLAVSE: \{\textit{GRAY}, \textit{16}, 5bits(i), 3bits(l)\}.}
	\label{tab:noise_snr_performance}
\end{table}

Fig. \ref{fig:noise_performance} details the results of two types of human-voiced noise, namely baby cry and background talkers. Under these types of noise, visual information becomes crucial in the SE task. Obviously, the AOSE(DP) method cannot give higher STOI scores than the “Noisy” speech, while all the AVSE methods outperform AOSE(DP). Even with the proposed compression units, LAVSE and iLAVSE still maintain acceptable performance in terms of both PESQ and STOI compared to AVDCNN. To further evaluate the proposed iLAVSE system on human-voiced noises at more SNR levels, we provide additional experimental results at mild SNR levels in Table \ref{tab:noise_snr_performance}. The results show that iLAVSE outperforms the AOSE(DP) baseline at all SNR levels.

\begin{figure}
	\centering
	\begin{subfigure}[!t]{0.47\linewidth}
		\centering
		\includegraphics[width=\linewidth, height=0.07\textheight]{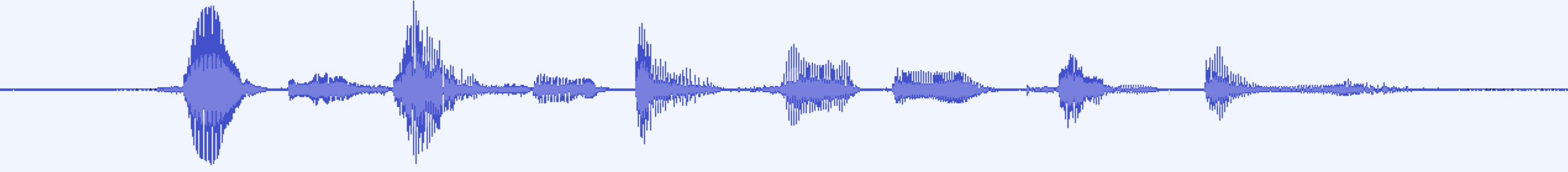}
		\caption{Clean waveform.}
	\end{subfigure}
	\begin{subfigure}[!t]{0.47\linewidth}
		\centering
		\includegraphics[width=\linewidth, height=0.07\textheight]{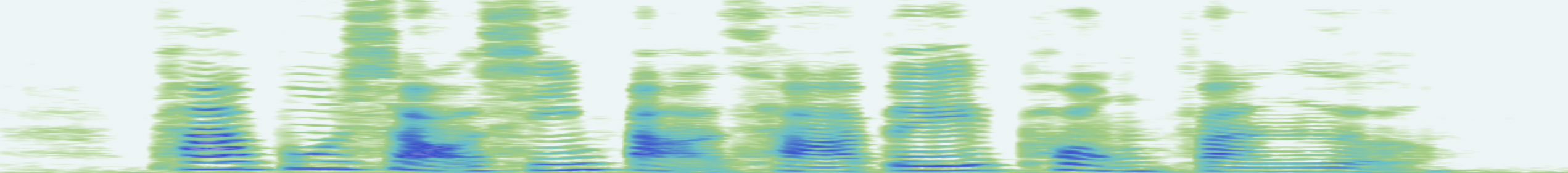}
		\caption{Clean spectrogram.}
	\end{subfigure}
	\par\medskip
	\begin{subfigure}[!t]{0.47\linewidth}
		\centering
		\includegraphics[width=\linewidth, height=0.07\textheight]{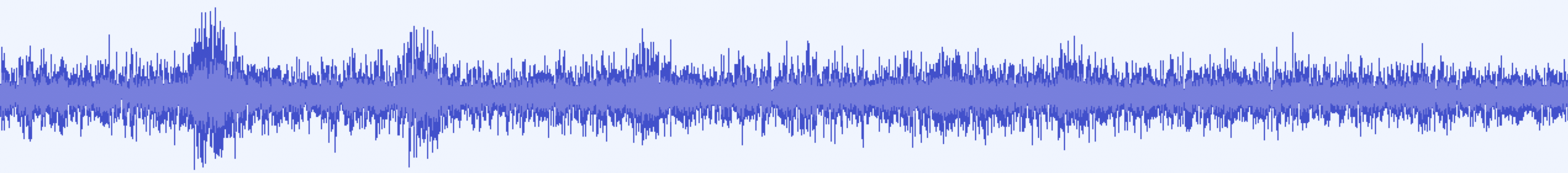}
		\caption{Noisy waveform.}
	\end{subfigure}
	\begin{subfigure}[!t]{0.47\linewidth}
		\centering
		\includegraphics[width=\linewidth, height=0.07\textheight]{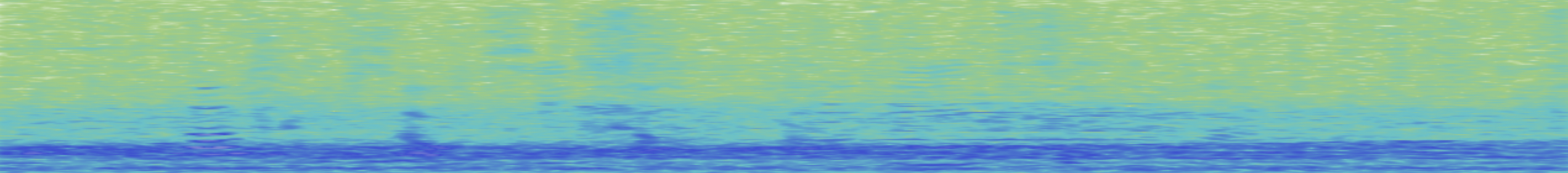}
		\caption{Noisy spectrogram.}
	\end{subfigure}
	\par\medskip
	\begin{subfigure}[!t]{0.47\linewidth}
		\centering
		\includegraphics[width=\linewidth, height=0.07\textheight]{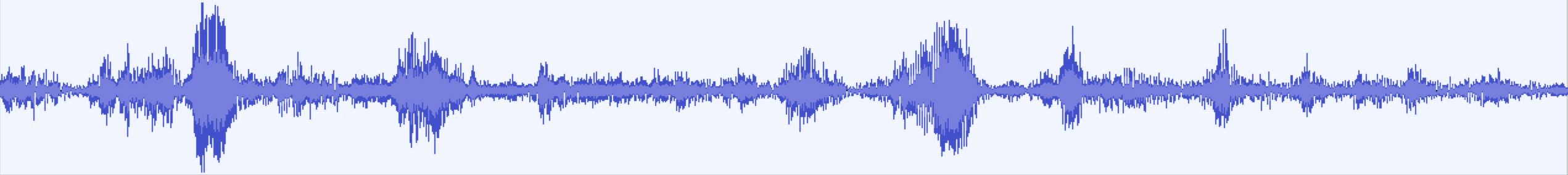}
		\caption{AOSE(DP) waveform.}
	\end{subfigure}
	\begin{subfigure}[!t]{0.47\linewidth}
		\centering
		\includegraphics[width=\linewidth, height=0.07\textheight]{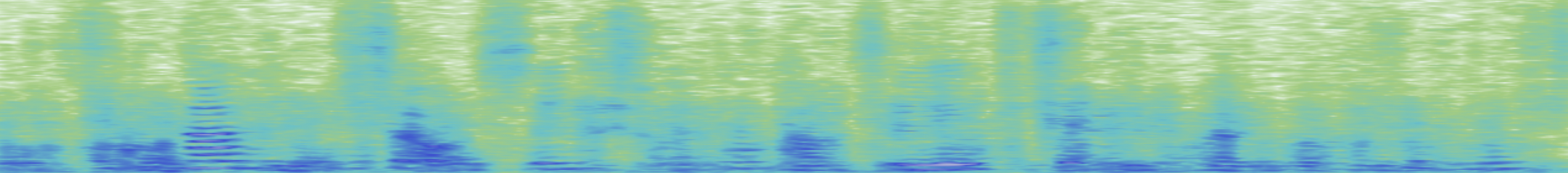}
		\caption{AOSE(DP) spectrogram.}
	\end{subfigure}
	\par\medskip
	\begin{subfigure}[!t]{0.47\linewidth}
		\centering
		\includegraphics[width=\linewidth, height=0.07\textheight]{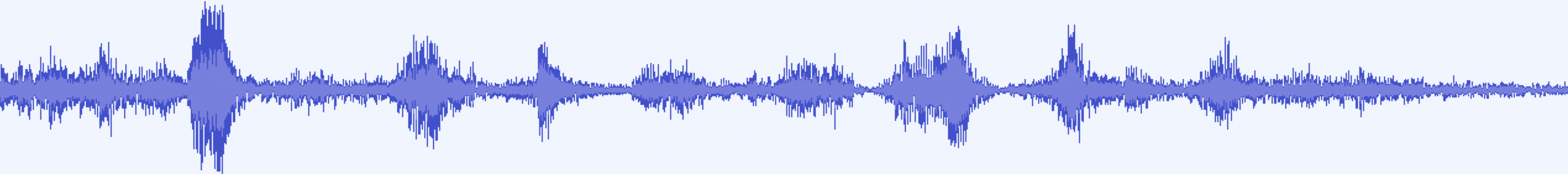}
		\caption{LAVSE waveform.}
	\end{subfigure}
	\begin{subfigure}[!t]{0.47\linewidth}
		\centering
		\includegraphics[width=\linewidth, height=0.07\textheight]{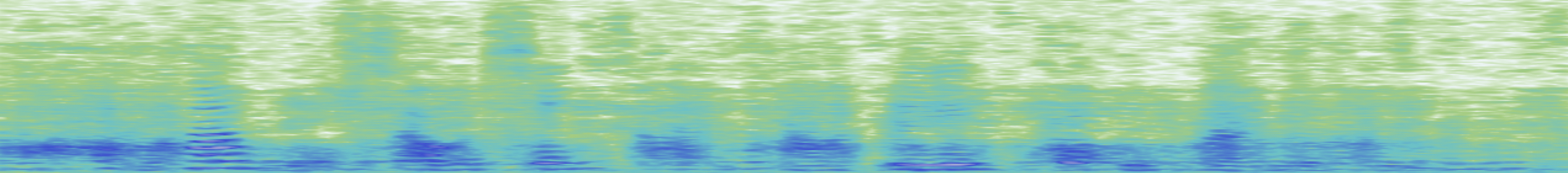}
		\caption{LAVSE spectrogram.}
	\end{subfigure}
	\par\medskip
	\begin{subfigure}[!t]{0.47\linewidth}
		\centering
		\includegraphics[width=\linewidth, height=0.07\textheight]{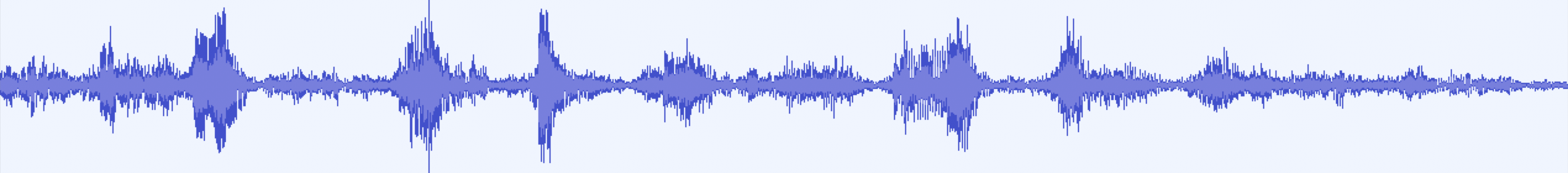}
		\caption{iLAVSE waveform.}
	\end{subfigure}
	\begin{subfigure}[!t]{0.47\linewidth}
		\centering
		\includegraphics[width=\linewidth, height=0.07\textheight]{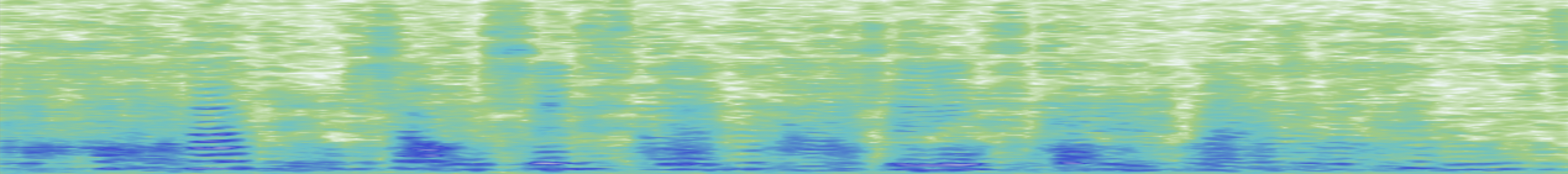}
		\caption{iLAVSE spectrogram.}
	\end{subfigure}
	\caption{The waveforms and spectrograms of an example speech utterance under the condition of street noise at {-7} dB. The vertical axis of the waveform figure represents the normalized amplitude (-0.1$\sim$0.1), and the vertical axis of the spectrogram figure represents the frequency (0k$\sim$8k Hz). The horizontal axis is time. The example utterance is 3 seconds long.}
	\label{fig:wav_spec}
\end{figure}

We further examined the spectrogram and waveform of the ``Noisy'' speech and the enhanced speech provided by AOSE(DP), LAVSE, and iLAVSE. An example under the condition of street noise at -7 dB is shown in Fig. \ref{fig:wav_spec}. The spectrogram and waveform of the clean speech are also plotted for comparison. From the figure, we see that iLAVSE can suppress the noise components in the noisy speech more effectively than AOSE(DP), and thus confirming the effectiveness of using the visual information. Furthermore, we note that the output plots of iLAVSE and LAVSE are very similar, which suggests that iLAVSE can still provide satisfactory performance even with compressed visual data.

\begin{figure}[t]
	\centering
	\includegraphics[width=1.0\linewidth]{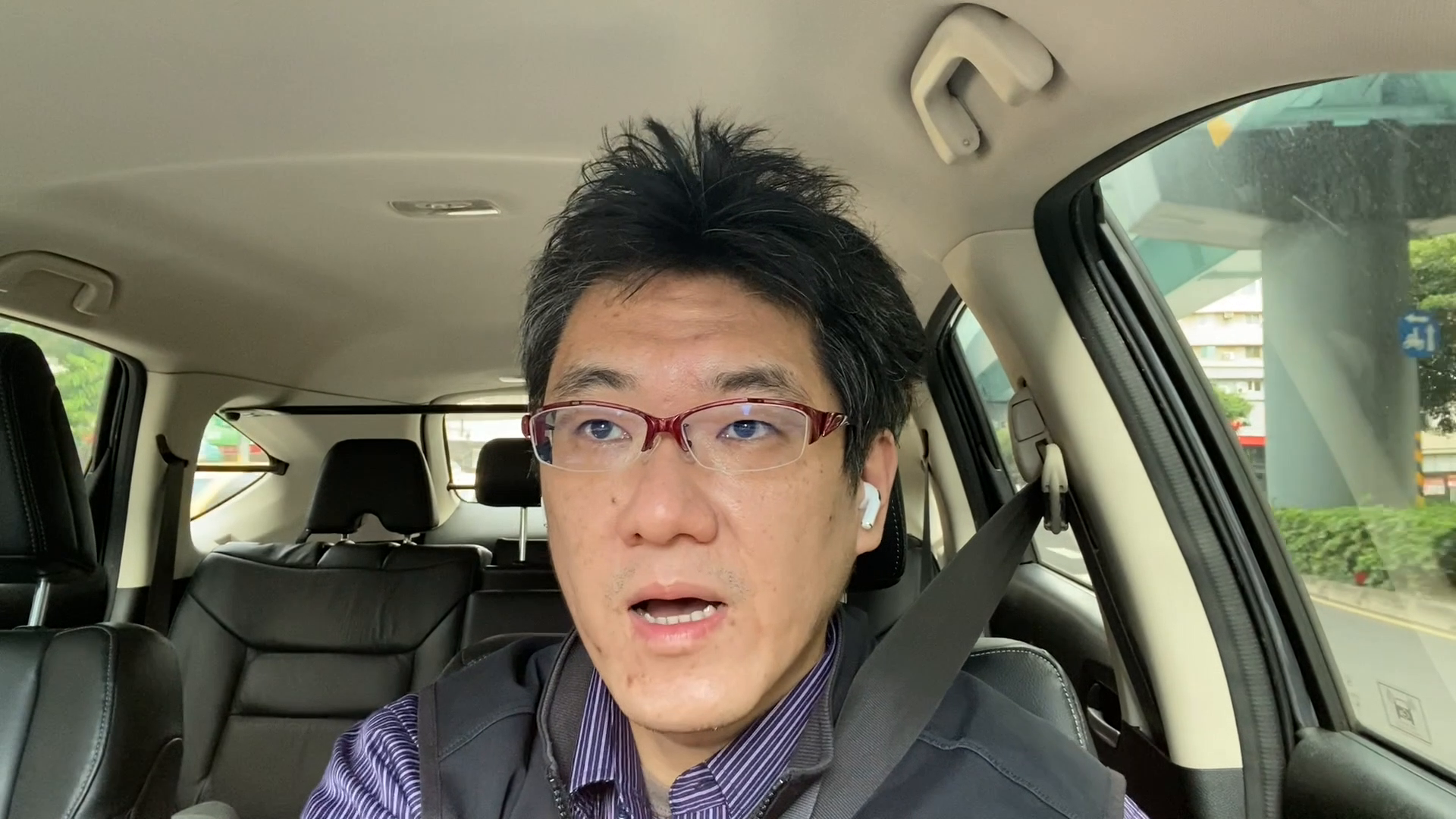}
	\caption{The real-world car-driving scenario.}
	\label{fig:frame}
\end{figure}

\begin{figure}
	\centering
	\begin{subfigure}[!t]{0.51\linewidth}
		\centering
		\includegraphics[width=\linewidth]{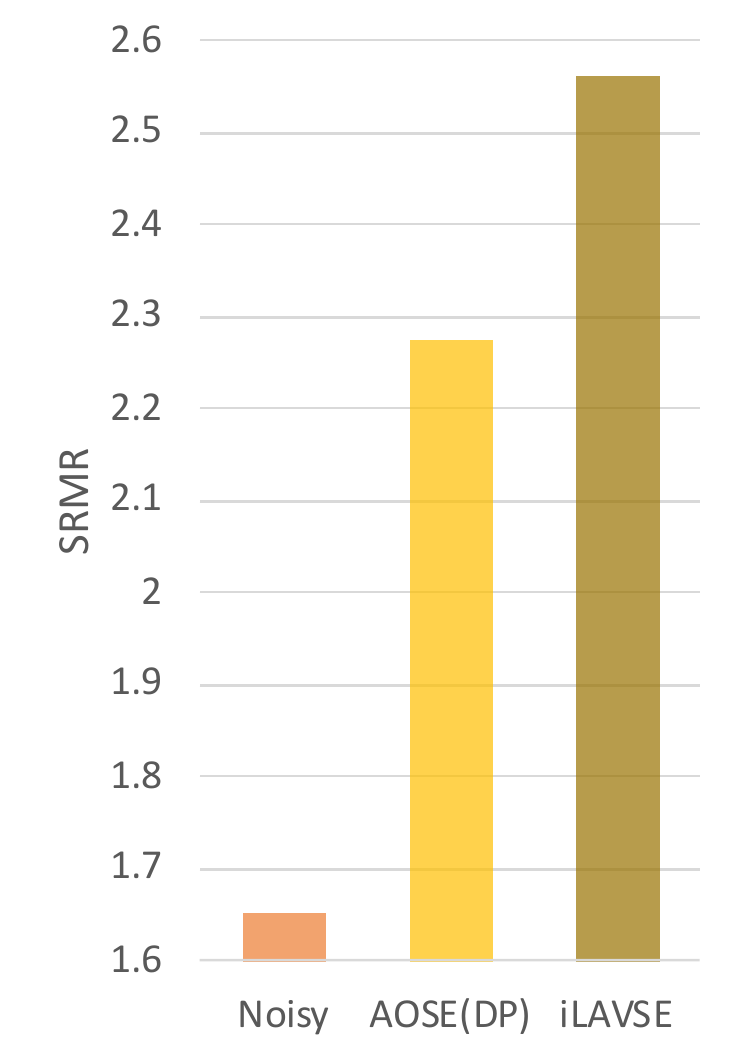}
		\caption{SRMR.}
		\label{fig:SRMR}
	\end{subfigure}
	\begin{subfigure}[!t]{0.46\linewidth}
		\centering
		\includegraphics[width=\linewidth]{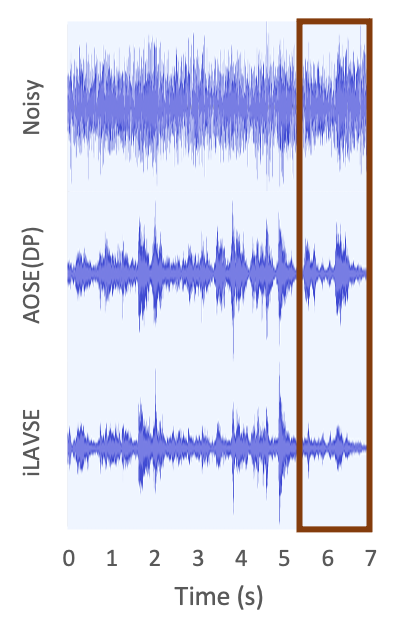}
		\caption{Waveforms.}
		\label{fig:real_wav}
	\end{subfigure}
	\caption{The average SRMR scores and sample processed waveforms obtained by AOSE(DP) and iLAVSE for the real-world videos. iLAVSE: \{\textit{GRAY}, \textit{16}, 5bits(i), 3bits(l)\}.}
	\label{fig:real}
\end{figure}

We recorded 10 video clips in a real car-driving scenario, as demonstrated in Fig. \ref{fig:frame}, with the background music and car-driving noise as our real-world testing data. The recording device was iPhone 12 Pro Max. Since there was no clean reference available in this set of experiments, we used the speech-to-reverberation modulation energy ratio (SRMR) \cite{falk2010non}, a non-intrusive modulation-spectral-representation-based metric for speech assessment to evaluate the performance of AOSE(DP) and iLAVSE. A higher SRMR score indicates better speech quality. The average SRMR scores and sample processed waveforms obtained by AOSE(DP) and iLAVSE for the real-world videos are shown in Fig. \ref{fig:SRMR} and Fig. \ref{fig:real_wav}, respectively. Fig. \ref{fig:SRMR} shows that the iLAVSE system achieves higher SRMR scores than the AOSE(DP) system and the original noisy speech. In Fig. \ref{fig:real_wav}, the top, middle, and bottom panels are the waveforms of the original noisy speech, AOSE-enhanced speech, and iLAVSE-enhanced speech, respectively. In the area framed by the brown box at the end of the speech, there is actually no speech, only background music. Obviously, the closed lips can effectively help iLAVSE to remove the background music, but the AOSE-enhanced speech still retains the background music.

\subsubsection{Asynchronization Compensation}

We simulated the audio-visual asynchronization condition by offsetting the visual and audio data streams of each utterance in the time domain. We designed 5 asynchronization conditions, i.e., 5 specific offset ranges (OFR): [-1, 1], [-2, 2], [-3, 3], [-4, 4], and [-5, 5]. For example, for OFR = [-1, 1], the offset range is from -1 to 1. An offset of -1, 0, or 1 frame (each frame = 20ms) was randomly selected (with equal probability) and used to shift the audio stream, so that the audio-visual asynchronization was -1, 0, or 1. In this way, we prepared 5 sets of training data with different degrees of audio-visual asynchronization. For the testing set, we simulated the audio-visual asynchronization condition using the fixed offsets in [-5, 5]. Therefore, the audio-visual data contained 11 different degrees of asynchronization. 

\begin{figure}
    \centering
    \begin{subfigure}[!t]{1.0\linewidth}
        \centering
        \includegraphics[width=\linewidth]{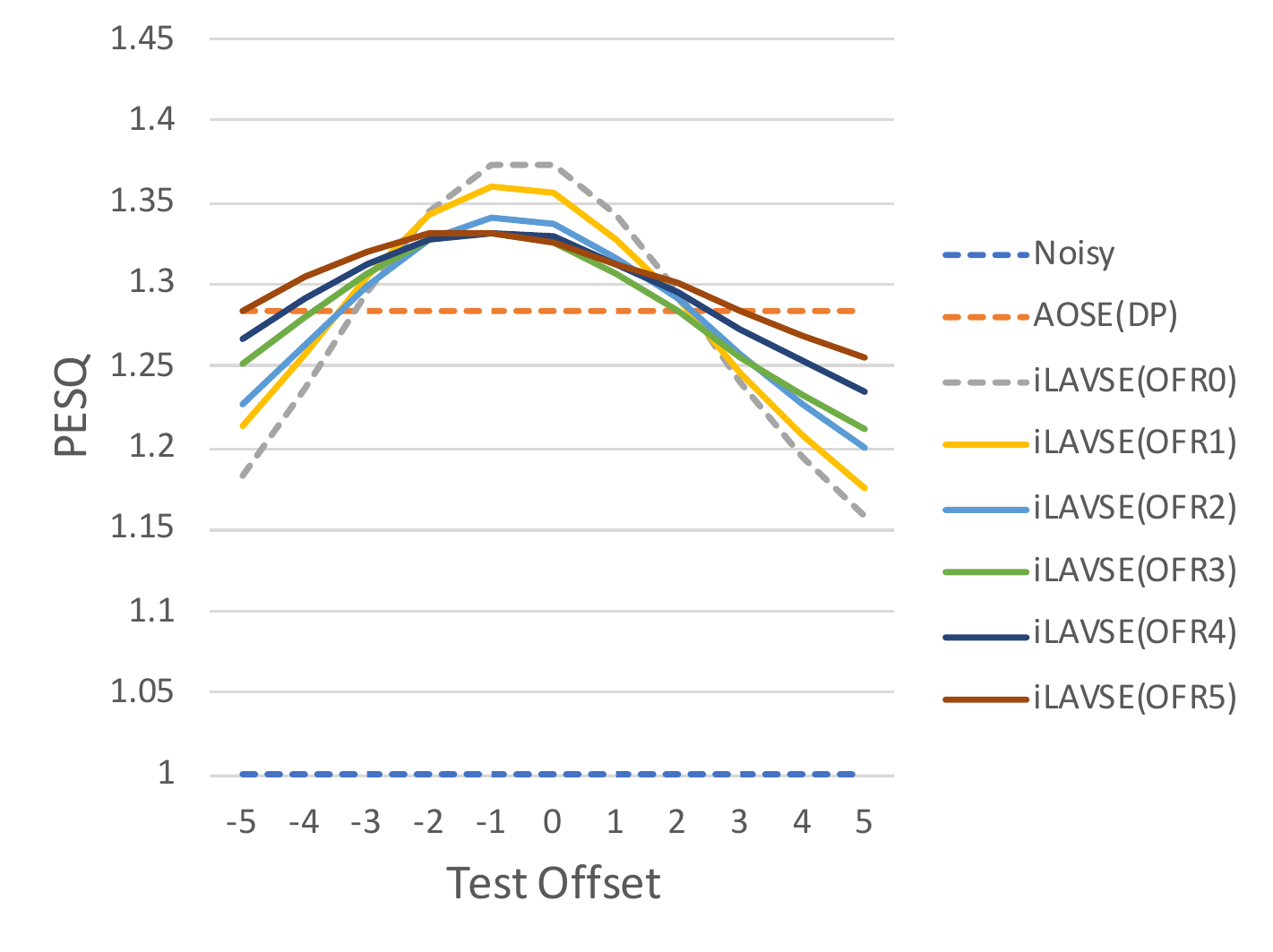}
        \caption{PESQ.}
        \label{fig:asynchronous_pesq}
    \end{subfigure}
    \par\medskip
    \begin{subfigure}[!t]{1.0\linewidth}
        \centering
        \includegraphics[width=\linewidth]{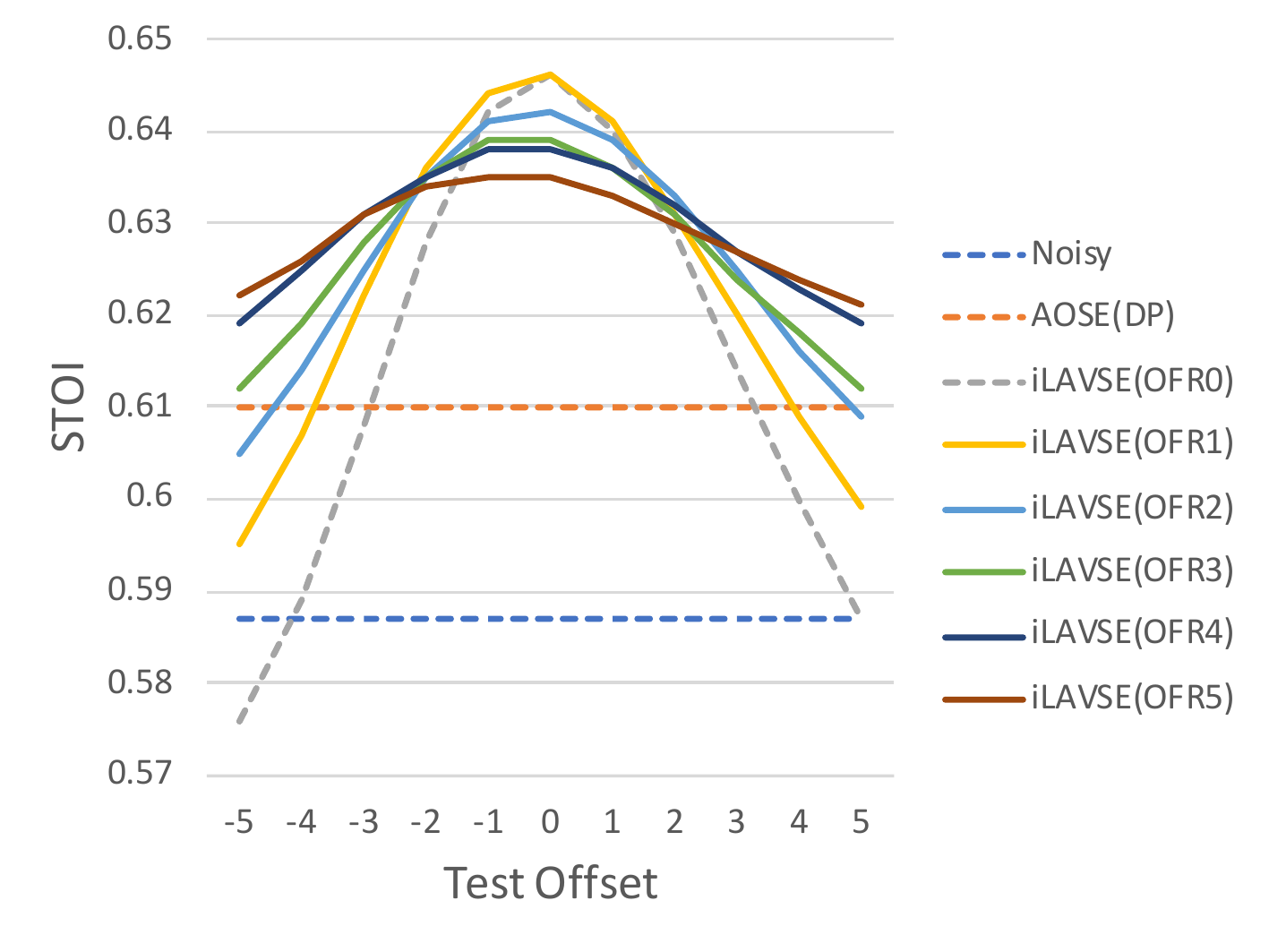}
        \caption{STOI.}
        \label{fig:asynchronous_stoi}
    \end{subfigure}
    \caption{The PESQ and STOI scores of iLAVSE trained and tested with different audio-visual asynchronous data.}
    \label{fig:asynchronous_score}
\end{figure}

Because the iLAVSE model was trained with 5 different OFRs, namely [-1, 1], [-2, 2], [-3, 3], [-4, 4], and [-5, 5], we therefore obtained 5 iLAVSE models, termed iLAVSE(OFR1), iLAVSE(OFR2), iLAVSE(OFR3), iLAVSE(OFR4), and iLAVSE(OFR5). These 5 models were then tested on the 11 different offsets (with a fixed offset in [-5, 5]). The results are shown in Fig. \ref{fig:asynchronous_score}. The results of Noisy, AOSE(DP), and iLAVSE trained without audio-visual asynchronization (denoted as iLAVSE(OFR0)) are also listed for comparison. All iLAVSE systems in this 
experiment used the original visual data.

Please note that, in both figures, the central point (cf. Test Offset = 0) represents the audio-visual synchronous condition. A ``Test Offset" value away from the central point indicates a more severe audio-visual asynchronous situation. ``Test Offset = -5" and ``Test Offset = 5" are the most severe conditions, where the audio and visual signals are misaligned for 5 frames (100 ms) in both cases. 

From Fig. \ref{fig:asynchronous_score}, we can note that when ``Test Offset = 0", iLAVSE(OFR0) achieves the best performance. This is reasonable because in this case, there is no asynchronous data in training and testing. When the asynchronization condition becomes severe, iLAVSE(OFR5) achieves better performance than other models. We also note that when the ``Test Offset" values lie in [-3, 3], iLAVSE(OFR5) always outperforms Noisy and AOSE(DP). The results confirm the effectiveness of including audio-visual asynchronous data (as augmented training data) to train the iLAVSE system to overcome the asynchronization issue. 

\subsubsection{Zero-Out Training}

We simulated the low-quality visual data condition by applying a low-quality percentage range (LPR) to the visual data. The low-quality percentage (LP) determines the percentage of missing frames in the visual data, and the LPR indicates the range of randomly assigned LPs for each batch. For example, if LPR is set to 10, LP will be randomly selected from 0\% to 10\%; if LP is set to 4\% for a batch with a length of 150 frames, a sequence of 6 ($150 \times 4\%$) frames of the visual data will be replaced with zeros. In this experiment, we chose LPRs $\in$ \{0, 10, 20, 30, 40, 50, 60, 70, 80, 90, 100\} for training, and set LPs $\in$ \{0, 10, 20, 30, 40, 50, 60, 70, 80, 90, 100\} to test the performance on specific percentages of missing visual data. The starting point of the missing visual part was randomly assigned for each batch.

\begin{figure}
    \centering
    \begin{subfigure}[!t]{1.0\linewidth}
        \centering
        \includegraphics[width=\linewidth]{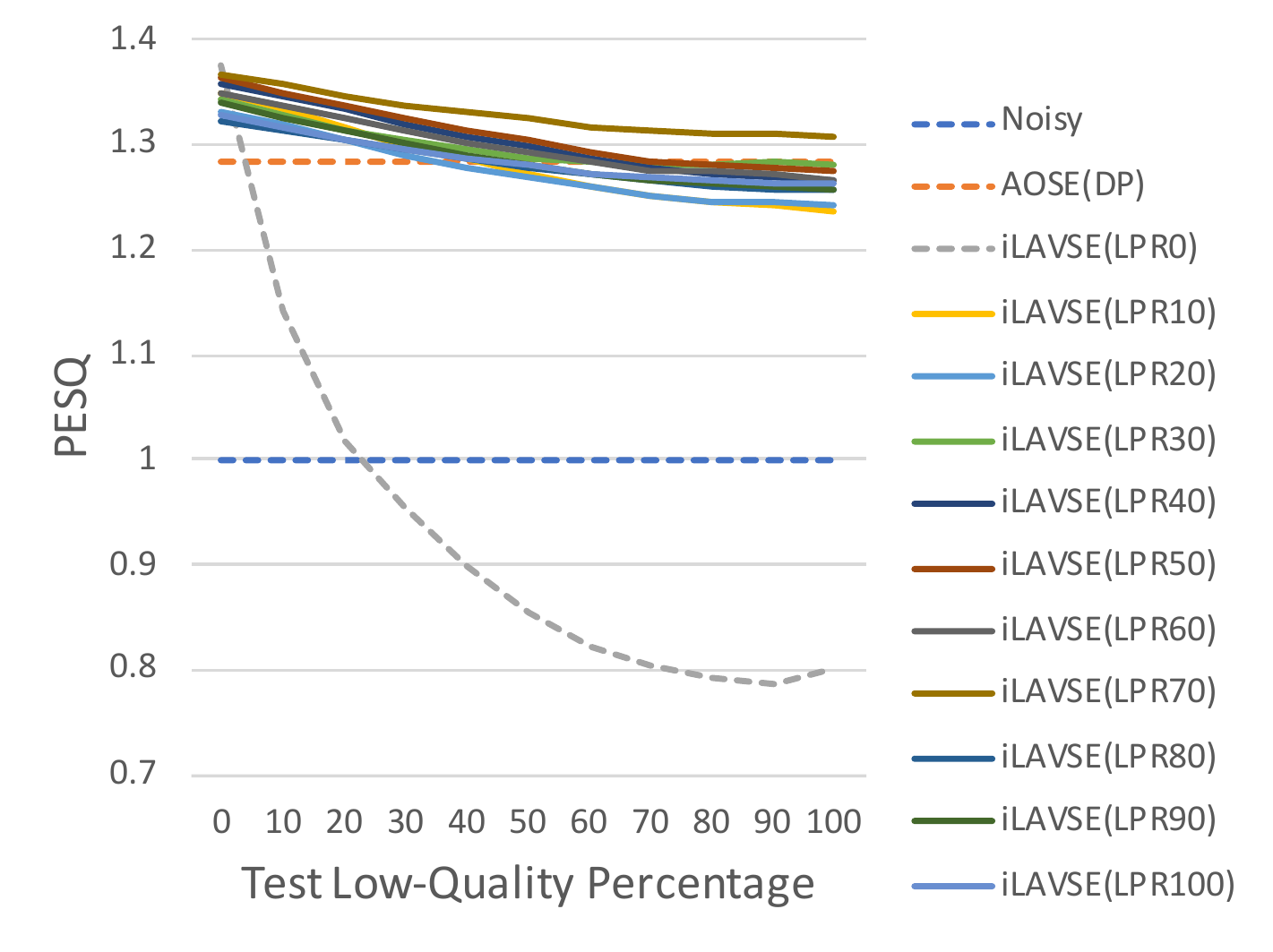}
        \caption{PESQ.}
    \end{subfigure}
    \par\medskip
    \begin{subfigure}[!t]{1.0\linewidth}
        \centering
        \includegraphics[width=\linewidth]{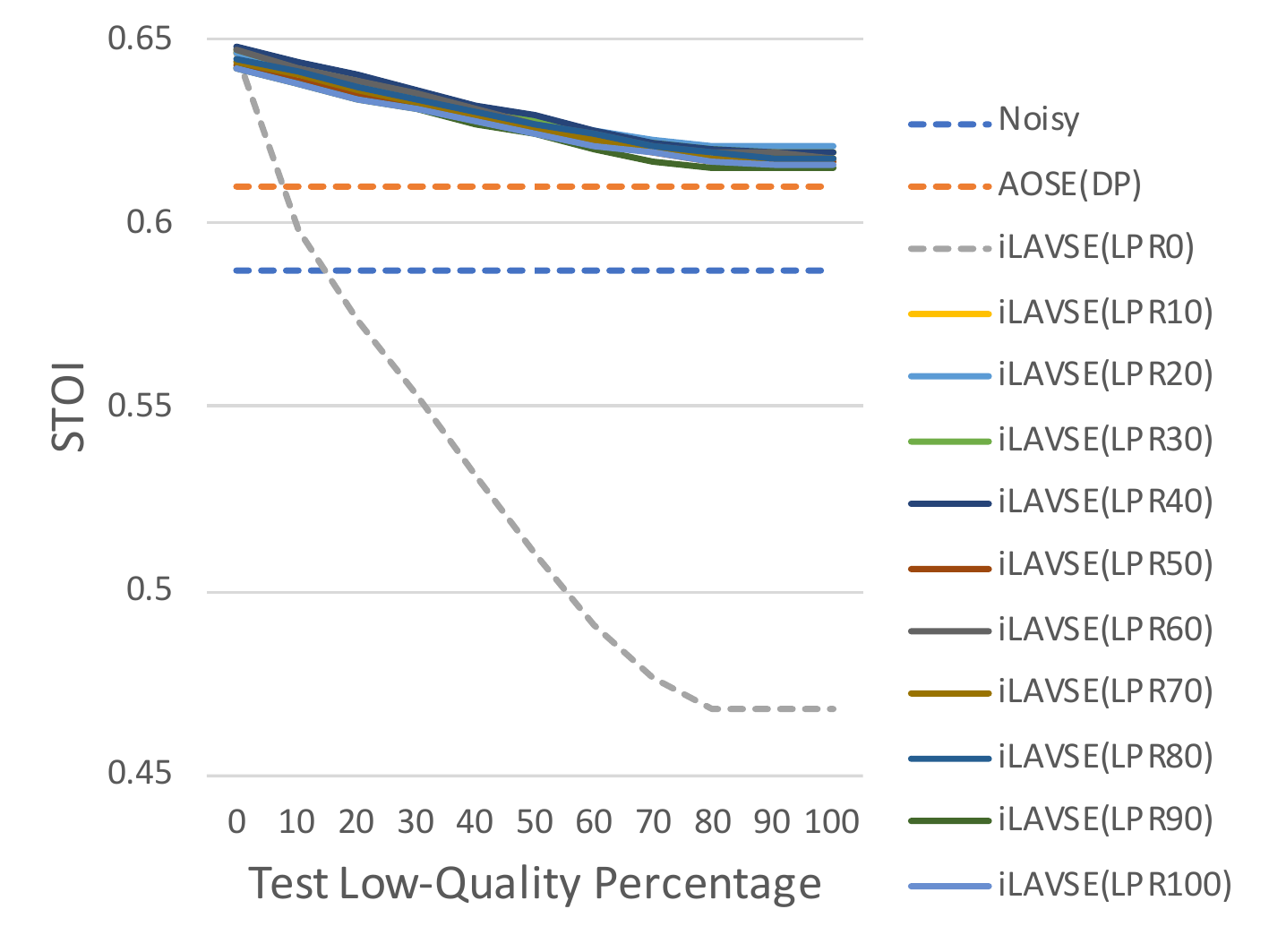}
        \caption{STOI.}
    \end{subfigure}
    \caption{The PESQ and STOI scores of iLAVSE trained with different LPRs and tested on specific LP conditions.}
    \label{fig:lq_score}
\end{figure}

The iLAVSE models trained with the 11 different LPRs are denoted as iLAVSE(LPR$i$), where $i$ = 0, ..., 10. The training set of iLAVSE(LPR0) did not contain missing visual data. A larger value of $i$ in LPR$i$ indicates a more severe low-quality visual data condition. The results are presented in Fig. \ref{fig:lq_score}, where the x-axis represents the LP value used for testing. The results in the figure show that without involving low-quality visual data in training (iLAVSE(LPR0)), the performance drops rapidly when visual data loss occurs in the testing data. The PESQ and STOI scores are even worse than those of Noisy and AOSE(DP). On the other hand, the iLAVSE models trained with low-quality visual data (even with low LPRs) are robust against all LP testing conditions. When the LP of the testing data is very high, the performance of iLAVSE converges to that of AOSE(DP), which shows that the benefit from visual information becomes negligible.

\section{Conclusion}
\label{sec:conclusion}

In this paper, we proposed the iLAVSE system, which aims to address three issues that may be encountered when developing practical AVSE systems, namely the high cost of processing visual data, audio-visual asynchronization, and low-quality visual data. The iLAVSE system includes three stages: data preprocessing, AVSE based on CRNN, and data reconstruction. The preprocessing stage uses the CRQ module and the AE module to extract the compact latent representation as the visual input of the AVSE stage. We used the data augmentation scheme and the zero-out training approach to solve the problems of audio-visual asynchronization and low-quality visual data, respectively. At present, due to the lack of relevant facilities, we cannot test the proposed model on a real low-resource computing platform. We can only compare the computing resources required by the new and old models and perform simulation experiments to verify our ideas. Our experimental results confirm that iLAVSE can effectively deal with these three practical issues and provide better SE performance than AOSE and related AVSE systems. Therefore, we believe that the proposed iLAVSE system is robust under adverse conditions and can be appropriately implemented in real-world applications.

In the present study, we focus on the application of the iLAVSE system in a car-driving scenario. In such a scenario, it is more common to encounter poor lighting issues than other adverse conditions, such as instance occlusion or noisy-image involvement, because a fixed camera can be used to directly monitor the driver’s face. In other application scenarios, we may use additional light sensors to signal the iLAVSE system when to use audio information alone. In the future, we will incorporate other neural network architectures, objective functions, and compression techniques \cite{puzicha2000spatial, patil2010edge, celebi2011improving} into the proposed system. In addition, we will further use the supplementary information provided by visual data, combined with self-supervised and meta learning, to improve the applicability of iLAVSE.

\section*{Acknowledgements}

This work was supported by the Ministry of Science and Technology [109-2221-E-001-016-, 109-2634-F-008-006-, 109-2218-E-011-010-, 110-2634-F-002-047] and Academia Sinica Career Development Award [AS-CDA-106-M04].

\bibliographystyle{IEEEbib}
\bibliography{refs}

\end{document}